\documentclass[aps,pre,twocolumn,groupedaddress]{revtex4-1}
\usepackage{ctable}
\usepackage[utf8]{inputenc}
\usepackage{amssymb}
\usepackage{amsmath}
\usepackage{bm}
\usepackage{graphicx}
\usepackage{xspace}
\usepackage[center]{subfigure}
\usepackage{wasysym}
\usepackage{multirow}
\usepackage{aas_macros}

\newcommand{\Ms}{\mathrm{M_s}}

\renewcommand{\Re}{\mathrm{Re}}
\newcommand{\Rm}{\mathrm{Rm}}

\newcommand{\apost}{\textit{a posteriori}\xspace}
\newcommand{\aprio}{\textit{a priori}\xspace}
\newcommand{\pd}[2]{\frac{\partial #1}{\partial #2}}
\newcommand{\flt}[1]{\overline{#1}}
\newcommand{\fav}[1]{\widetilde{#1}}

\newcommand{\Enzo}{\textsc{Enzo}\xspace}

\renewcommand{\v}[1]{\boldsymbol{ #1}}
\newcommand{\bottomrulea}[1]{\left( #1 \right)}
\newcommand{\grad}[1]{\nabla \bottomrulea{#1}}
\newcommand{\curl}[1]{\nabla \times #1}
	
\renewcommand{\div}[1]{\nabla \cdot \bottomrulea{#1}}

\newcommand{\EMFdata}{{\bm{\mathcal{E}}}}
\newcommand{\rres}{\overline{\rho}}
\newcommand{\jres}{\overline{\mathbf{J}}}
\newcommand{\Jflt}{\jres}

\newcommand{\pres}{\overline{P}}

\newcommand{\EMF}{\widehat{\bm{\mathcal{E}}}}
\newcommand{\tuijdata}[1][]{\tau_{ij}^{\mathrm{u} #1}}
\newcommand{\tbijdata}[1][]{\tau_{ij}^{\mathrm{b} #1}}

\newcommand{\Ekinsgsdata}{E^{\mathrm{u}}_{\mathrm{sgs}}}
\newcommand{\Emagsgsdata}{E^{\mathrm{b}}_{\mathrm{sgs}}}
\newcommand{\Esgsdata}{E_{\mathrm{sgs}}}

\newcommand{\EkinsgsSStar}{\widehat{E}^{\mathrm{u},\mathcal{S}^*}_{\mathrm{sgs}}}

\newcommand{\EmagsgsM}{\widehat{E}^{\mathrm{b},\mathcal{M}}_{\mathrm{sgs}}}

\newcommand{\nuk}{\nu^\mathrm{u}}

\renewcommand{\SS}{\mathrm{SS}}
\newcommand{\EV}{\mathrm{EV}}

\newcommand{\NL}{\mathrm{NL}}

\newcommand{\tu}[1][]{\widehat{\tau}_{ij}^{\mathrm{u} #1}}
\newcommand{\tb}[1][]{\widehat{\tau}_{ij}^{\mathrm{b} #1}}

\newcommand{\Sflt}[1][]{\fav{\mathcal{S}}}
\newcommand{\Mflt}[1][]{\flt{\mathcal{M}}}

\newcommand{\pb}{\beta_{\mathrm{p}}}

\newcommand{\abs}[1]{\left | #1 \right |}

\newcommand{\bra}[1]{\left( #1 \right)}

\DeclareMathOperator{\Skew}{skew}
\DeclareMathOperator{\Kurt}{kurt}
\newcommand{\Do}{\textsc{GS}}
\newcommand{\Dt}{\textsc{F3}}
\newcommand{\Df}{\textsc{F5}}

\makeatletter
\def\widebreve#1{\mathop{\vbox{\m@th\ialign{##\crcr\noalign{\kern3\p@}%
      \brevefill\crcr\noalign{\kern3\p@\nointerlineskip}%
      $\hfil\displaystyle{#1}\hfil$\crcr}}}\limits}

\def\brevefill{$\m@th \setbox\z@\hbox{$\braceld$}%
  \bracelu\leaders\vrule \@height\ht\z@ \@depth\z@\hfill\braceru$}
\makeatletter

\begin{document}

\title{Comparative statistics of selected subgrid-scale models in large eddy 
simulations of decaying, supersonic MHD turbulence}

\author{Philipp Grete}
\email{\footnotesize{grete@pa.msu.edu}}
\affiliation{Max-Planck-Institut f\"ur Sonnensystemforschung, Justus-von-Liebig-Weg 3,
D-37077 G\"ottingen, Germany}
\affiliation{Institut f\"ur Astrophysik, Universit\"at G\"ottingen,
	  Friedrich-Hund-Platz 1, D-37077 G\"ottingen, Germany}

\author{Dimitar G Vlaykov}
\affiliation{Max-Planck-Institut f\"ur Dynamik und Selbstorganisation, 
Am Faßberg 17, D-37077 G\"ottingen, Germany}

\author{Wolfram Schmidt}
\affiliation{Hamburger Sternwarte, Universit\"at Hamburg, Gojenbergsweg 112, 
D-21029 Hamburg, Germany}

\author{Dominik R G Schleicher}
\affiliation{Departamento de Astronom\'ia, Facultad Ciencias F\'isicas y Matem\'aticas, 
Universidad de Concepci\'on, Av. Esteban Iturra s/n Barrio Universitario, Casilla 160-C, Chile}

\date{\today}

\begin{abstract}
Large eddy simulations (LES) are a powerful tool in understanding processes that 
are inaccessible by direct simulations due to their complexity, for example, in
the highly turbulent regime.
However, their accuracy and success depends on a proper subgrid-scale (SGS) model 
that accounts for the unresolved scales in the simulation.
We evaluate the applicability of two traditional SGS models, namely the  eddy-viscosity (EV) and 
the scale-similarity (SS) model, and one recently proposed nonlinear (NL) SGS model in the realm
of compressible MHD turbulence.
Using 209 simulations of decaying, supersonic (initial sonic Mach number
$\Ms\approx3$) MHD turbulence with a shock-capturing scheme and varying
resolution, SGS model and filter,
we analyze the ensemble statistics of kinetic and magnetic energy spectra and 
structure functions.
Furthermore, we compare the temporal evolution of lower and higher order statistical moments 
of the spatial distributions of kinetic and magnetic energy, vorticity, 
current density, and dilatation magnitudes.
We find no statistical influence on the evolution of the flow by any model if 
grid-scale quantities are used to calculate SGS contributions.
In addition, the SS models, which employ an explicit filter, have no impact in general.
On the contrary, both EV and NL models change the statistics if an explicit filter is used.
For example, they slightly increase the dissipation on the smallest scales.
We demonstrate that the nonlinear model improves higher order statistics already
with a small explicit filter, i.e. a three-point stencil.
The results of e.g. the structure functions or the skewness and kurtosis of the
current density distribution are closer to the ones obtained from simulations at 
higher resolution.
In addition, no additional regularization to stabilize the model is required.
We conclude that the nonlinear model with a small explicit filter is suitable for 
application in more complex
scenarios when higher order statistics are important.
\end{abstract}


\maketitle

\section{Introduction}
Magnetohydrodynamic (MHD) turbulence is observed in many different processes
and on many different scales,
for example, in astrophysics, in magnetized accretion disks \citep{RevModPhys.70.1},
stellar winds \citep{1995ARA&A..33..283G}, galaxies and galaxy mergers
\cite{Rodenbeck2016}, or more generally in processes like
magnetic reconnection \citep{2015ApJ...806L..12O} and the turbulent
amplification of magnetic fields \citep{PhysRevE.92.023010}.
Moreover, experiments on Earth also aim at a better understanding of e.g.
flow-driven MHD instabilities \citep{2014PhPl...21a3505C}.
However, the full multi-dimensional dynamics are only rarely accessible in 
these observations and experiments.
For this reason, simulations are frequently used as a third, complementary 
approach or to support the design of experiments \citep{Tzeferacos201524}.

Simulations of pure turbulence are nowadays possible at very
high resolution \cite{Kaneda2003,Yeung13102015,Federrath2016} and properly 
capture a lot of physical processes.
However, there are still many extreme regimes, for example
in astrophysics, where turbulence is thought to play an important role but which
are inaccessible to direct simulations with realistic parameters.
This situation is also not going to change in the near future
despite the ever increasing availability and performance of large computing 
clusters, and the advances in numerical methods.
In these cases, large eddy simulations (LES) have been employed successfully in the
past, however, mostly in the realm of (incompressible) hydrodynamics, see e.g. 
\citep{Sagaut2006} for a general introduction and \citep{lrca-2015-2} for an 
astrophysics related review.
In LES only the largest eddies, which correspond to motions on large and 
intermediate scales, are simulated directly.
The smallest scales, which are either not represented or unresolved in these
simulations, are reintroduced by the means of a subgrid-scale (SGS) model.
In other words, LES try to incorporate effects on the large scale flow that 
stem from the small scales or from interaction between large and small scales.
Ideally, the overall quality of the simulation improves with respect to the 
physical processes that are taken into account.

Formally, this is equivalent to applying a low-pass filter to the ideal 
compressible MHD equations resulting in expressions
of the form \citep{Vlaykov2016a}
\begin{eqnarray}
\label{eq:masscons}
  \pd{\rres}{t}+\div {\rres \fav{\v{u}}} = 0,\\
\label{eq:momcons}
  \begin{split}
  \pd{\flt{\rho} \fav{\v{u}}}{t} 
  + \div{\flt{\rho} \fav{\v{u}} \otimes \fav{\v{u}} 
    - \flt{\v{B}} \otimes \flt{\v{B}}} 
     + \grad{\pres + \frac{\flt{B}^2}{2}}  \\*
    = 
    - \nabla \cdot \tau, 
    \end{split} \\
\label{eq:fluxcons}
    \pd{\flt{\v{B}}}{t} - \curl\bra{\fav{\v{u}} \times \flt{\v{B}}} =
    \curl{\EMFdata},
\end{eqnarray}
for a static, homogeneous and isotropic filter under periodic boundary 
conditions.
The filtered primary quantities, i.e. the density $\flt{\rho}$, velocity
$\fav{\v{u}}$, magnetic field $\flt{\v{B}}$ (incorporating $1/\sqrt{4\pi}$)
and pressure $\flt{P}$ are considered resolved in LES.
Normal filtering is denoted by an overbar $\flt{\Box}$ whereas mass-weighted,
Favre \citep{1983PhFl...26.2851F} filtering is denoted by a tilde
$\fav{\Box} = \flt{\rho\Box}/\flt{\rho}$.

Assuming an isothermal equation of state ($\flt{P} \propto \flt{\rho}$), 
all interactions between resolved and unresolved scales and among unresolved
scales themselves are captured by the two new terms in the equations.
The turbulent stress tensor is given by
\begin{eqnarray}
  \label{eq:tau_def}\tau_{ij} =  \tuijdata - \tbijdata +
  \bra{\flt{B^2} - \flt{B}^2}\frac{\delta_{ij}}{2} \;,
\end{eqnarray}
and can be decomposed into the turbulent (or SGS) magnetic pressure 
(last term), SGS Reynolds stress $\tuijdata$ and SGS Maxwell stress $\tbijdata$ with
\begin{eqnarray}
  \tuijdata \equiv \flt{\rho} \bra{\fav{u_i u_j} - \fav{u}_i \fav{u}_j}
  \;\; \text{and} \;\;
  \tbijdata \equiv \bra{\flt{B_i B_j} - \flt{B}_i~\flt{B}_j} . \label{eq:tudata}
\end{eqnarray}
The second new term is the turbulent electromotive force (EMF):
\begin{eqnarray}
  \EMFdata = \flt{\v{u}\times\v{B}} - \fav{\v{u}} \times \flt{\v{B}}\label{eq:EMFdata} \;.
\end{eqnarray}
Moreover, the total filtered energy is given by
\begin{eqnarray}
	\label{eq:SGSen}
	\flt{E} = \underbrace{\frac{1}{2} \flt{\rho} \fav{u}^2 + \frac{1}{2}\flt{B}^2}_
	{\mathrm{(resolved)}}
	+ \underbrace{\frac{1}{2}\flt{\rho} \bra{\fav{u^2} - \fav{u}^2} + \frac{1}{2} \bra{\flt{B^2} - \flt{B}^2}}_
	{= \Ekinsgsdata + \Emagsgsdata \equiv \Esgsdata \mathrm{(unresolved)}}
\end{eqnarray}
where, by virtue of the identity $\tau_{ii}^\Box \equiv 2 \Esgsdata^\Box$ (with
Einstein summation), the
SGS energies are given by the traces of the corresponding stress tensor.

All these SGS terms are unclosed because the mixed terms, i.e. $\fav{u_i u_j}$,
$\flt{B_i B_j}$ and $\flt{\v{u}\times\v{B}}$, are not explicitly accessible in an
LES and require modeling.
This modeling is the main challenge for a successful LES.

SGS models have been subject of research in hydrodynamics for 
many decades, e.g. in the incompressible \citep{Sagaut2006}, 
compressible \citep{2009lesc.book.....G} and astrophysical \citep{lrca-2015-2} 
regime.
However, work in the realm of MHD and in particular compressible MHD is
scarce, see \citep{Chernyshov2014} and \citep{Miesch2015} for recent reviews.
Directly linked to this work are the MHD simulations of (decaying) turbulent boxes
in 3D \citep{Miki2008}, in 2D \citep{1994PhPl....1.3016T} and 
in the incompressible case \citep{Mueller2002,Yokoi2013}.
However, all these groups use different numerical schemes, e.g. 
finite-differences or (pseudo-) spectral methods, while we employ a
shock-capturing finite-volume scheme.
Usually, these shock-capturing methods are thought to provide an
implicit SGS model as shocks are captured by locally increasing the
effective numerical dissipation with the help of e.g. slope limiting.
Since this procedure is part of the overall method, these simulations are
also referred to as implicit LES (ILES) \cite{grinstein2007implicit}.

In the presented work, we compare the \apost behavior of several SGS model 
including a nonlinear model that explicitly captures 
compressibility \cite{Vlaykov2016a}.
Their performance was previously evaluated \aprio \citep{1367-2630-17-2-023070,Grete2016a}.
Here, we analyze a set of simulations of 
decaying homogeneous and isotropic turbulence with respect to a set 
of statistical quantities including energy spectra, structure functions 
and statistical moments of the primary fields. 
More details on the numerics and the implementation are given in the following
section \ref{sec:methods} where we also introduce the different models tested 
and the setup of the particular test case of decaying compressible MHD turbulence.
Afterwards, in section \ref{sec:results} we present the results of different 
statistics such as energy spectra, evolution of statistical moments and structure
function.
Then we discuss these results with respect to previous work in \ref{sec:discussion}
and conclude in the last section \ref{sec:conclusions}.

\section{Method}
\label{sec:methods}
\subsection{Subgrid-scale models}
In our \aprio analysis \citep{Grete2016a} we tested three different model families,
eddy-dissipation, scale-similarity, and nonlinear models, with different normalizations.
All models were tested against the expressions \eqref{eq:tudata} and \eqref{eq:EMFdata}
where the filtered nonlinear term is not split into additional components.
A split allows to separate different types of interactions between scales: 
only among resolved scales, between resolved and unresolved scales, and only among
unresolved scales.
Thus, the previous \aprio results evaluated the performance of the models to
recover all kind of interactions simultaneously.
This is also what we implicitly expect of the models in this work.
For each model family we identified the best model with corresponding coefficients.
These models are now tested \apost and described in the following.

\paragraph{The eddy-viscosity} ($\EV$) model has the longest tradition with
roots going back even further than its formulation for LES by
Smagorinsky \citep{Smagorinsky1963}.
While originally developed for the kinetic SGS stress tensor in hydrodynamics, 
the general idea has been transferred to MHD \citep{Yokoi2013,Miki2008}, where the EMF 
closure is usually referred to as anomalous or eddy-resistivity.
The names of these functional models stem from their primary feature: purely
dissipative behavior analogous to e.g. molecular viscosity and resistivity.
We use the following formulations
\begin{eqnarray}
	\label{eq:tu}
	\tu &=& -2  \flt{\rho} \nuk \mathcal{\fav{S}}^*_{ij} 
	+ 2/3 \delta_{ij} \EkinsgsSStar 
	\;, \\
	\label{eq:tb}
	\EMF &=& -  \eta_t \Jflt  \;,
\end{eqnarray}
with the resolved traceless kinetic rate-of-strain tensor 
    $\mathcal{\fav{S}}^*_{ij}=1/2\bra{\fav{u}_{i,j}+\fav{u}_{j,i}}-1/3\delta_{ij}\fav{u}_{k,k}$
    and current density $\Jflt=\nabla\times\flt{\v{B}}$. 
The strengths of the eddy-viscosity $\nuk$ and eddy-resistivity $\eta_t$
are given by
\begin{eqnarray}
    \nuk &=& C_{1} \Delta \sqrt{\EkinsgsSStar/\flt{\rho}} \quad \text{and} \\
	\eta_t &=& C_{2} \Delta \sqrt{\bra{\EkinsgsSStar + \EmagsgsM}/\flt{\rho}} \;.
\end{eqnarray}
They are locally scaled by the SGS energies derived from realizability 
conditions of the SGS stresses resulting in \citep{FLM:353319,1367-2630-17-2-023070}
\begin{eqnarray}
\label{eq:EnClosures}
\EkinsgsSStar &=& C_3 \Delta^2 \flt{\rho} |\Sflt^*|^2  \quad
\text{and} \\
\EmagsgsM &=& C_4 \Delta^2 |\Mflt|^2 
\end{eqnarray}
with 
$\Mflt_{ij}=1/2\bra{\flt{B}_{i,j}+\flt{B}_{j,i}}$
being the resolved magnetic rate-of-strain tensor.
In agreement with the \aprio analysis \citep{Grete2016a} the coefficients are 
set to $C_1 = C_2 = 0.05$ and $C_3 = C_4 = 0.04$.
Please note that the coefficient indices are different from the ones in the
referenced paper due to a reduced selection of models.
Furthermore, the SGS Maxwell stress is neglected as the eddy-diffusivity extension
in compressible MHD was found to not match the reference data (correlations $<0.1$)
in the \aprio analysis.
Therefore, the effects of the SGS Maxwell stress are modeled implicitly by the 
numerical scheme.

\paragraph{The scale-similarity} ($\SS$) model originates in experimental
hydrodynamics where it was observed that scale-to-scale energy
transfer is self-similar up to intermittency \citep{BARDINA1980}.
Formally, this additional scale separation can be expressed by an additional
(test) filter $\widebreve{\Box}$ whose filter width is larger than the 
original one. 
The model is given by
\begin{eqnarray}
\tu &=& 
C_{5} \widebreve{\flt{\rho}} \bra{\widebreve{\fav{u_i}\fav{u_j}} 
- \widebreve{\fav{u_i}}\widebreve{\fav{u_j}}} \;, \\
\tb &=& 
C_{6} \bra{\widebreve{\flt{B_i}\flt{B_j}} 
- \widebreve{\flt{B_i}}\widebreve{\flt{B_j}}} \;, \\
\EMF &=&  
C_{7} \bra{\widebreve{\fav{\v{u}}\times\flt{\v{B}}} - 
\widebreve{\fav{\v{u}}}\times\widebreve{\flt{\v{B}}}}
\;,
\end{eqnarray}
and mass-weighted filtering also applies to the test filter where
velocity components are involved.
Again, we choose the coefficients according to the \aprio analysis:
$C_5 = 0.67$, $C_6 = 0.9$ and $C_7 = 0.89$.
The model allows for energy transfer down- and up-scale and, as a
structural closure, aims at reproducing closely the properties of the SGS terms
and not just their effects on the large scales.

\paragraph{The nonlinear} ($\NL$) model is another structural model and
exhibited the highest correlations with reference data in \aprio tests
\citep{1367-2630-17-2-023070,Grete2016a}.
It can be derived from Taylor expansion of the inverse filter kernel
\citep{Yeo87,Vlaykov2016a} and requires no further assumptions about the
underlying flow features.
We employ the primary compressible extension resulting in the following
model:
\begin{eqnarray}
\tu &=& 
\frac{1}{12} \Delta^2 \flt{\rho} \fav{u}_{i,k} \fav{u}_{j,k} \;, \\
\tb &=& 
\frac{1}{12} \Delta^2  \flt{B}_{i,k} \flt{B}_{j,k} \;, \\
\EMF &=&  
 \frac{1}{12} \Delta^2 \varepsilon_{ijk}  \bigl( \fav{u}_{j,l} \flt{B}_{k,l} 
 - \bra{\ln \flt{\rho}}_{,l} \fav{u}_{j,l} \flt{B}_{k} \bigr) 
\;.
\end{eqnarray}
As previously shown, this model does not require 
a calibration coefficient \aprio \cite{Grete2016a} and the
prefactor $1/12$ originates from the second moment of the Gaussian or 
box filter.

\subsection{Implementation and explicit filtering}
We implemented the different models in the open-source, community-developed
magnetohydrodynamic code \textsc{Enzo} \citep{Enzo2013}.
The new terms are handled by operator-splitting within the MUSCL-Hancock framework 
and evaluated with centered finite-differences.
They are advanced in time together with the other fluxes by the existing second order 
Runge-Kutta scheme.

Furthermore, we implemented a flexible filtering framework that supports
different stencils and weights in real space.
In order to determine the individual weights, we construct discrete, 
explicit filters 
by minimizing the residual between analytic and discrete filter yielding
so called optimal filters \citep{Vasilyev199882}.
We optimize for wavenumbers below the filter width \citep{FLD:FLD914}.
The resulting weights and filter widths for a symmetric one-dimensional 
3-point (\Dt), and 5-point (\Df) stencil of
the box kernel are listed in table~\ref{tab:filterWeights}.
\begin{table}
\caption{Filter weights for a discrete representation of the box filter based on an
optimal filter approach by minimizing the residual between the analytic and discrete filter
over wavenumbers below the filter width.
The filter width $\flt{\Delta}$ is given in terms of grid-spacing $\Delta_x$.}
\label{tab:filterWeights}       
\begin{ruledtabular}
\begin{tabular}{lcccc}
\multirow{2}{*}{Identifier} & filter width& \multicolumn{3}{c}{filter weights}\\
& $\flt{\Delta}$ & $w_i$ &  $w_{i \pm 1}$ &  $w_{i\pm2}$  \\
\noalign{\smallskip}\hline\noalign{\smallskip}
\Do & $1\Delta_x$ & $1$ & $0$ & $0$ \\
\Dt & $2.711\Delta_x$ & $0.4015$ & $0.29925$ & $0$ \\
\Df & $4.7498\Delta_x$ & $0.20238$ & $0.22208$ & $0.17673$ 
\end{tabular}
\end{ruledtabular}
\end{table}
We construct the corresponding multi-dimensional filters of 
$N$-point one-dimensional stencils by 
\begin{equation}
	\flt{\Box}_{i,j,k} = 
	\sum_{l=-N^*}^{N^*} 
	\sum_{m=-N^*}^{N^*} 
	\sum_{n=-N^*}^{N^*} 
	w_l w_m w_n \Box_{i+l,j+m,k+n}
	\;,
\end{equation}
with discrete filter weights $w_i$, $N^* = (N-1)/2$, 
and indices referring to discrete spatial locations.
This translates to the sequential application of the one-dimensional filters 
and results in large 3-d filter stencils ($N^3)$, e.g. 27 points for the 
\Dt~filter and 125 points for the \Df~filter.
However, this construction is more accurate 
\citep{FLD:FLD914}
than the alternative approach of simultaneous application.
Finally, we also use the trivial grid filter (\Do).
In that case, the quantities are used as they are computed in the original
simulation, which
corresponds to a natural filter by the discretization itself.

\subsection{Simulations}
All our simulations are run with \textsc{Enzo} \citep{Enzo2013} using the HLL Riemann 
solver within the MUSCL-Hancock framework with second order Runge-Kutta time 
integration.
Moreover, we close the equations of ideal MHD with a quasi-isothermal equation 
of state, i.e. we employ an ideal equation of state with a ratio of specific
heats $\kappa = 1.001$. 

In order to get proper initial conditions for freely decaying, compressible
MHD turbulence, we first start from uniform initial conditions 
$\rho =  1$, $\v{u} = \v{0}$ and $\v{B} = \bra{0.6325,0,0}^\mathrm{T}$ 
(in code units) corresponding to an
initial plasma beta (ratio of thermal to magnetic pressure) of $\pb = 5$.
These uniform initial conditions are then driven in a cubic box with length $L=1$ and
resolution $512^3$ by a stochastic forcing field
that evolves in space and time \citep{Schmidt2009}.
The forcing field has a parabolic profile between wavenumber $1 < k < 3$ 
and is centered at $k=2$.
Moreover, the overall forcing amplitude is set to $V = 3$ and distributed 
between $1/3$ compressive and $2/3$ solenoidal components.
The forcing leads to statistically isotropic, homogeneous turbulence with
root mean square sonic Mach number of $\Ms \approx 3$ after
two turnover times $T = L/(2V)$ and we follow its evolution for a total of 
$20T$.

Assuming that two different snapshots are statistically independent from each
other after one turnover time, we have an ensemble of 19 different realizations
(at $t=\{2,3,\dots,20\}T$)
of isotropic, homogeneous turbulence, which are statistically indistinguishable.
We take states from this ensemble 
as initial conditions for freely decaying
turbulence.
This later allows us to analyze ensemble statistics to better capture
the intermittent nature of turbulence.

For each realization we run the following 11 simulations with different 
configurations -- varying  the resolution, models (or lack thereof) 
and explicit filter, namely:  
\begin{itemize}
	\item 3 implicit large eddy simulations (ILES). Recall that in these simulations
		there is no explicit model ($\tau = \EMFdata = 0$), at resolutions
		of $128^3$, $256^3$ and $512^3$.
		They are referred to as ILES-128, ILES-256, and ILES-512, 
		respectively.
	\item 3 LES with the eddy-viscosity model at a resolution of $128^3$:  
        one with no explicit filter (EV-\Do); 
        one with a filter with a three-point stencil (EV-\Dt); 
        and one with a five-point stencil (EV-\Df).
	\item 2 LES with the scale-similarity model at $128^3$ employing three- 
		(SS-\Dt) and five-point (SS-\Df) explicit filtering (because a grid-scale
		scale-similarity model does not exist).
	\item 3 LES with the nonlinear model again at $128^3$ with all three different filters
		NL-\Do, NL-\Dt, and NL-\Df.
\end{itemize}
The highest resolution $512^3$ ILES simulations are later used as reference runs.
Comparing the results between the different ILES enables us to evaluate the pure
influence of reduced resolution (and thus reduced dynamics) on the evolution of 
the decay.
For example, the Reynolds number that can be achieved in a simulation depends
on the resolution.
Given that we use a shock-capturing scheme to solve the ideal MHD equations, 
i.e.~viscosity and resistivity are 
of numerical nature rather than explicit,
the kinetic and magnetic Reynolds numbers are not readily accessible.
In ILES of decaying, homogeneous isotropic turbulence 
the effective kinetic Reynolds number can be estimated as~\cite{PhysRevE.89.013303}
\begin{align}
	\Re = - \left < \abs{\nabla \times \v{u}}^2 \right > L^2 
	\bra{\frac{\mathrm{d} \v{u}^2}{\mathrm{d}t}}^{-1} \;.
\end{align}
Extending this concept to obtain an estimate of the magnetic Reynolds number
yields
\begin{align}
	\Rm = - \left < \abs{\nabla \times \v{B}}^2 \right > L^2 
	\bra{\frac{\mathrm{d} \v{B}^2}{\mathrm{d}t}}^{-1} \;.
\end{align}
\begin{table}
\caption{Effective initial kinetic and magnetic Reynolds numbers in the simulations
depending on resolution. 
The numbers are estimated according to~\cite{PhysRevE.89.013303}.}
\label{tab:Reynolds}       
\begin{ruledtabular}
\begin{tabular}{lcc}
Resolution & $\Re$ & $\Rm$ \\
\noalign{\smallskip}\hline\noalign{\smallskip}
$128^3$ & $\mathcal{O}\bra{600}$ &  $\mathcal{O}\bra{450}$ \\
$256^3$ & $\mathcal{O}\bra{1100}$ &  $\mathcal{O}\bra{900}$ \\
$512^3$ & $\mathcal{O}\bra{1700}$ &  $\mathcal{O}\bra{1400}$
\end{tabular}
\end{ruledtabular}
\end{table}
The resulting initial effective Reynolds numbers for the individual 
resolutions are listed in table~\ref{tab:Reynolds}.
Thus, the effective magnetic Prandtl number of $\mathcal{O}\bra{1}$.
Finally, we can draw conclusions on the influence of the individual 
models on the evolution from a comparison of results between ILES and 
the different LES at identical resolution (here $128^3$) 

Each simulation follows the decay for two turnover times.
We capture snapshots every $0.05T$ resulting in 40 snapshots per simulation.
Finally, we note that the initial conditions at lower resolutions ($128^3$ and 
$256^3$) have been calculated from the initial $512^3$ snapshot of each 
realization by coarse-graining, i.e. volume-averaging over $2^3$ and 
$4^3$ cells, respectively.
We choose this approach to minimize the differences in the initial
conditions between the individual configurations of a particular
realization.
In addition to this, all simulations, including the LES, first decay for $0.2T$ 
without model before the actual $2T$ decay that we follow and analyze.
This is done in order to obtain converged spectra at a given resolution, because
the process of coarse-graining leaves excess energy at the smallest scales and
the interaction between model and excess energy is unknown.
Moreover, resolution dependent quantities, e.g. magnetic energy or vorticity 
(see next section), relax to their intrinsic value in this transient-decay.

\section{Results}
\label{sec:results}

\subsection{Energy spectra}

\begin{figure*}[htbp]
	\begin{center}
        \includegraphics[width=170mm]{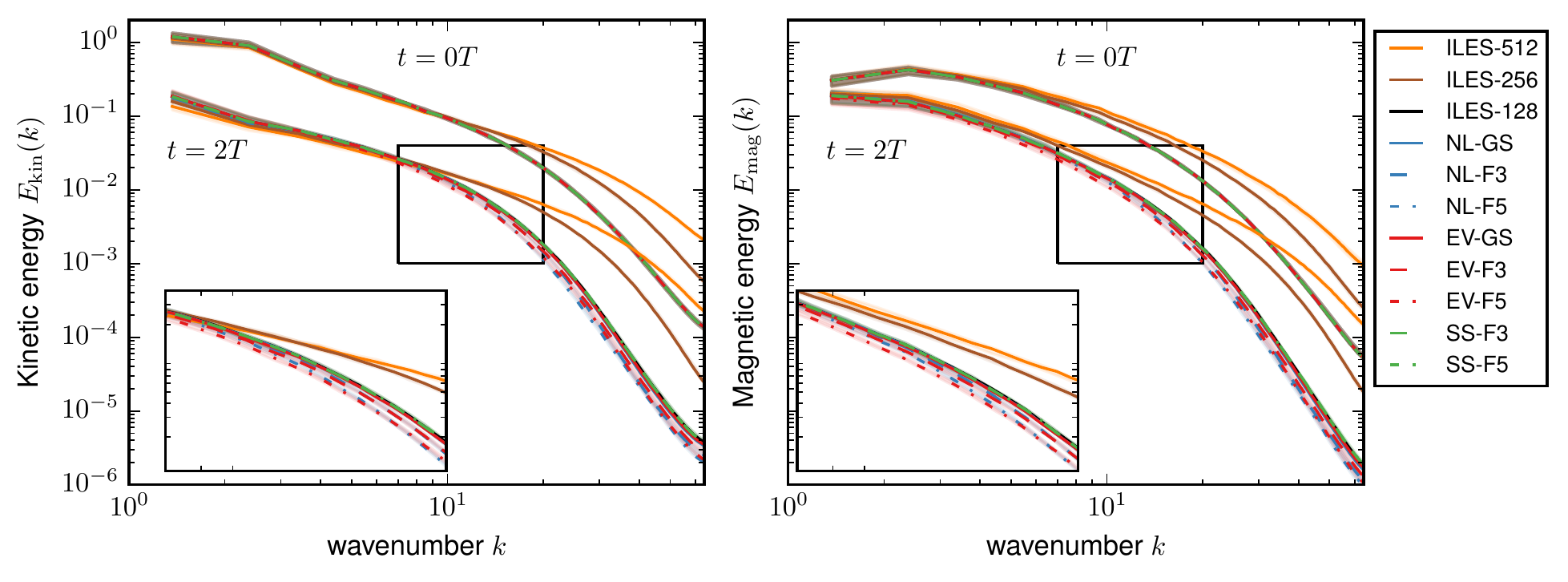}
	\end{center}
	\caption{Kinetic (left) and magnetic (right) 
	energy spectra at initial time $t=0T$ and 
		after two turnover times $t=2T$ of free decay.
		The lines correspond to the median over all 19 realizations and the shaded
		areas (if not hidden by the linewidth) show the interquartile range.
		The kinetic energy spectrum has been calculated based on the
		Fourier transform of $\sqrt{\rho}\v{u}$.
        The insets show a magnification of the configurations in the 
        intermediate wavenumber range at $t=2T$.
	}
	\label{fig:EnSpec}
\end{figure*}
Figure~\ref{fig:EnSpec} shows the kinetic and magnetic energy spectra 
initially at $t=0T$ and after two turnover times at $t=2T$.

Initially, there is basically no variation of the individual spectra 
between different configurations.
However, the difference in resolution is clearly visible.
The highest resolution ensemble (ILES-512) exhibits the most extended power-law 
range in the kinetic energy spectrum.
Accordingly the wavenumber $k$ where the spectrum drops 
due to numerical dissipation is shifted towards larger scales
with decreasing resolution (ILES-256 and ILES-128).
This also confirms our approach of removing coarse-graining artifacts in the 
initial snapshots by the initial transient-decay.
We verified that the smallest scales 
are statistically stationary in the following evolution.
The small peak still visible around $k=2$ is due to the original characteristic 
driving scale of the initial forced simulation.

After two turnover times, the differences between resolutions remain the most 
striking feature in the spectra with respect to the wavenumbers where the spectrum
wears off.
At the lowest resolution, the differences between the ILES-128 and the different SGS 
models are subtle.
There exists virtually no difference between the ILES-128 and the grid-filtered
eddy-viscosity (EV-GS) and nonlinear (NL-GS) model or the scale-similarity runs
(SS-F3 and SS-F5) - both in the kinetic and in the magnetic spectrum.
The SGS runs of the eddy-viscosity model with explicit filtering lead to
a slightly stronger dissipative behavior. 
This is expressed by a marginal reduction of energy on the smallest scales 
$k \gtrsim 30$ and more pronounced ($\approx40\%$) for the larger filter width
(EV-F5) than for the smaller filter width ($\approx20\%$ - EV-F3).
The nonlinear model (NL-F3 and NL-F5) seems to have a very similar dissipative 
behavior to the eddy-viscosity model given that the spectra coincide for the
same filter.

Finally, the simulations are still approximately isotropic after the two
turnover times as measured via the generalized Shebalin angle $\theta$ 
\cite{shebalin1983,oughton1994} defined by
\begin{align}
    \tan^2 \theta = \frac{\sum k_\perp^2 E(k)}{\sum k_x^2 E(k)},
\end{align}
with $k_\perp^2 = k_y^2 + k_z^2$.
A fully isotropic spectrum yields $\theta=\tan^{-1}\sqrt{2}\approx54.7^{\circ}$,
whereas a fully anisotropic spectrum with all energy in the perpendicular
modes yields $\theta=90^\circ$.
Here, for all simulations, i.e. independent of resolution, SGS model and filter,
the generalized Shebalin angle changes from 
$\theta\approx(55\pm1)^\circ$ at $t=0T$ to 
$\theta\approx(57\pm1)^\circ$  at $t=2T$ 
for both the kinetic and magnetic
energy spectrum.
Thus, the weak mean field is not expected to have a significant influence on the flow
over the free decay.

\subsection{Temporal evolution of mean quantities}
\label{sec:resMeans}
\begin{figure}[htbp]
	\begin{center}
		\includegraphics[width=85mm]{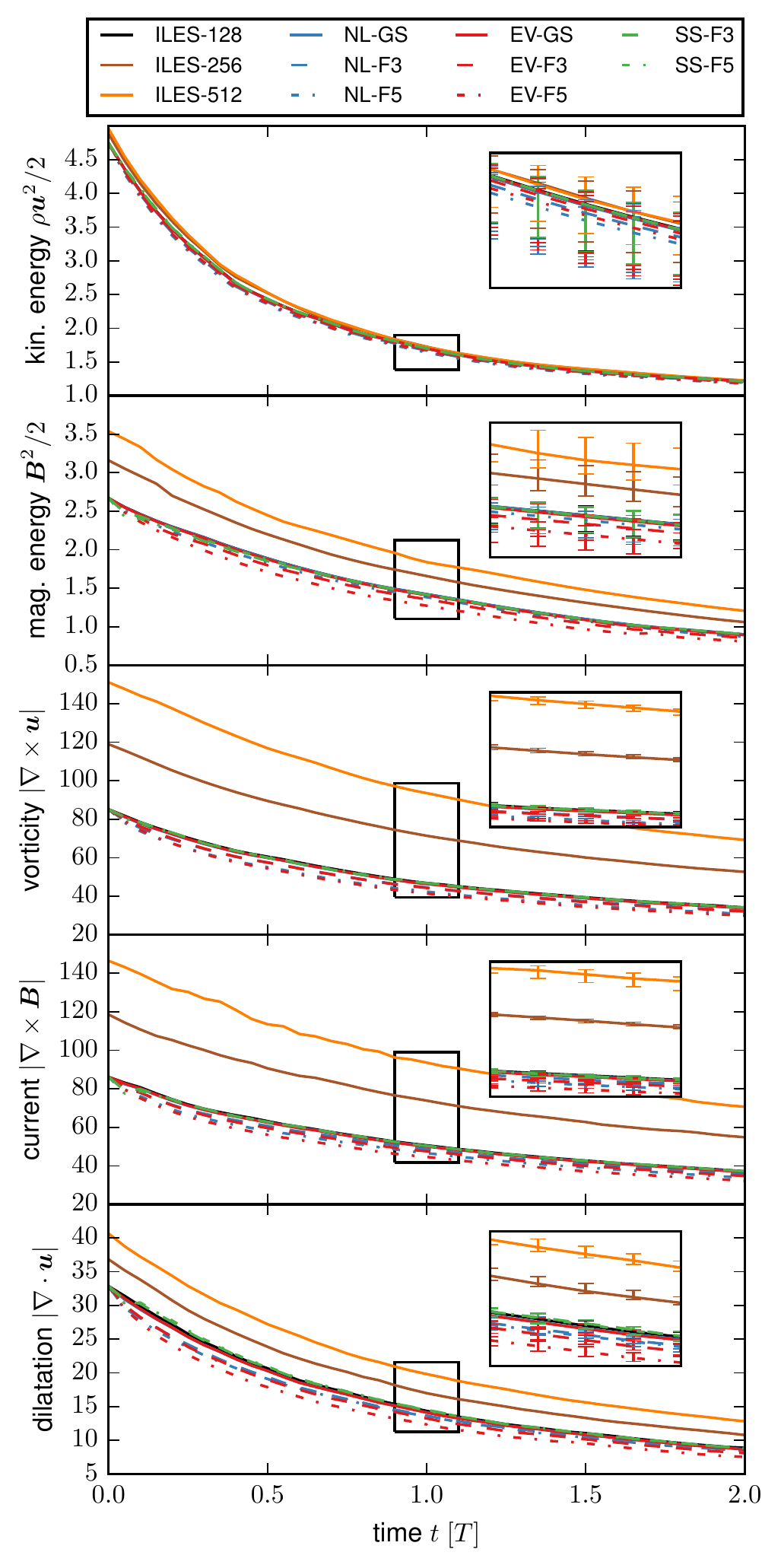}
	\end{center}
	\caption{Temporal evolution of the ensemble median (over 19 different
	realizations) of the spatial mean kinetic energy, magnetic energy,
	vorticity magnitude, current density magnitude and dilatation magnitude.
	The variations in terms of interquartile ranges are illustrated in
	the insets.
	}
	\label{fig:EvolMeans}
\end{figure}
The evolution over time of the spatially averaged kinetic energy, magnetic energy, 
vorticity magnitude,
current density magnitude and dilatation magnitude is shown in 
Fig.~\ref{fig:EvolMeans}.
Overall, all quantities smoothly decay over the two turnover times as
expected.
The panels show a similar behavior of the SGS models as observed in the energy 
spectra.
However, there are subtle differences.

The evolution of the kinetic energy, $\rho\v{u}^2/2$, is converged with
respect to resolution and SGS model.
In contrast to this, the magnetic energy, $\v{B}^2/2$, shows a clear 
separation with resolution.
The decreased turbulence intensity or effective Reynolds number at lower 
resolutions cannot sustain the original magnetic field strength of the
driven simulation conducted at higher resolution.
Thus, the differences in the initial values at $t=0T$ can be attributed to
the removal of coarse-graining artifacts (here, the excess magnetic energy
for a given resolution) by the transient-decay.
When removing the resolution effects, e.g. by normalizing each ensemble to its
initial value, all configurations but one (EV-F5) collapse to a converged
evolution.
The eddy-viscosity model with the largest explicit filter shows a
$\approx 10\%$ decrease in magnetic energy indicating increased dissipation.
However, in contrast to the energy spectra, here not only the small scales
are affected by the model, but a back-reaction onto the largest scales has
taken place.

The derived quantities, i.e. the vorticity magnitude $\abs{\curl{\v{u}}}$, the
current density magnitude $\abs{\curl{\v{B}}}$ and the dilatation magnitude
$\abs{\nabla \cdot \v{u}}$, exhibit comparable behavior.
In the raw data, resolution effects dominate and 
resolving smaller spatial scales leads to larger values.
Contrary to the evolution of the magnetic energy, this effect does not vanish
when all configurations are normalized and a lower resolution results in a slightly 
increased decay rate.
The same four SGS models (EV-F3, EV-F5, NL-F3 and NL-F5)
as in the energy spectra  now separate from the bulk 
(EV-GS, NL-GS, SS-F3 and SS-F5), which is
indistinguishable from the ILES-128 ensemble.
The vorticity and current magnitudes are reduced by $5\%$ (NL-F3),
$6\%$ (EV-F3), $10\%$ (NL-F5) and $12\%$ (EV-F5) indicating
very similar behavior with respect to filter width between eddy-viscosity and nonlinear 
model.
In contrast, the dilatation magnitude is only reduced by $5\%$ (F3) and $6\%$ (F5) for
the NL models, but by $8\%$ (F3) and $16\%$ (F5) for the EV models.
The latter indicates that the eddy-viscosity model is more isotropic than the
nonlinear one in agreement with their functional form.

\subsection{Higher-order statistics}
After having described the evolution of mean quantities, we now consider the
temporal evolution of the higher order moments of the distributions.
They provide insight into the tails of the distributions, which are crucial in 
the characterization and understanding of, for instance, the intermittency of turbulence.
In general, the variances of all quantities (kinetic and magnetic energy, and
vorticity, current and dilatation magnitudes) posses the same characteristics 
as their means in the previous section, i.e. an overall decay is observed 
with similar ensemble variations and configuration separations.

\begin{figure*}[htbp]
	\begin{center}
		\includegraphics{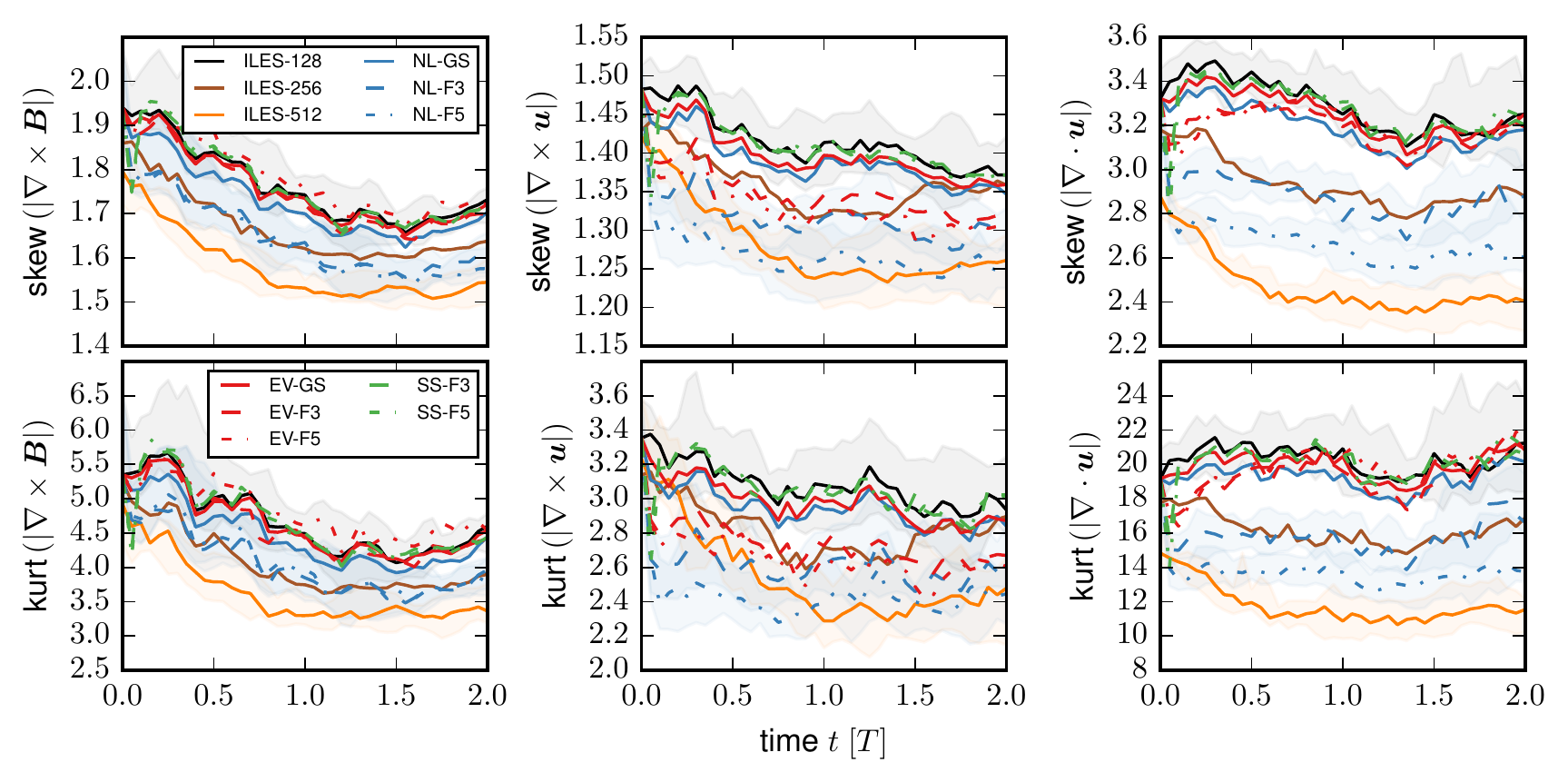}
	\end{center}
	\caption{Temporal evolution of the skewness and kurtosis of the distributions
		of current density (left column), vorticity (center column)
		and dilatation (right column)  magnitude. 
	The lines indicate the median over all 19 realizations.
	The shaded areas correspond to the interquartile ranges (IQR).
	For clarity, they are only shown for ILES-128, ILES-512, NL-F3 and NL-F5 as
	the IQRs of similar lines are virtually identical, e.g. the lines of
    ILES-128, EV-GS, NL-GS, SS-F3 and SS-F5, 
}
	\label{fig:EvolHigherOrder}
\end{figure*}

The next higher order moments we consider are the skewness
\begin{equation}
	\Skew{x} = \frac{\left < \bra{x - \left < x \right > }^3 \right > }{\sigma^3 \bra{x}}  \;,
\end{equation}
with standard deviation $\sigma$ and the (Fisher) kurtosis
\begin{equation}
	\Kurt{x} = \frac{\left < \bra{x - \left < x \right > }^4 \right > }{\sigma^4 \bra{x}} -3 \;. 
\end{equation}
The kinetic and magnetic energy skewness and kurtosis do not discriminate 
between the different models as the ensemble variations for each 
configuration are larger than the differences between the configurations.
This picture changes when looking at the higher order moments of the derived quantities.
The temporal evolution of the skewness and kurtosis of the vorticity, dilatation
and current magnitude are very similar (with respect to the qualitative 
evolution of the medians and interquartile ranges) as shown in 
Fig.~\ref{fig:EvolHigherOrder}.
Thus, we focus on the magnitude of the current density for a quantitative discussion.
Firstly, both skewness and kurtosis are resolution dependent.
A lower resolution increases the skewness by $\approx  5\%$ (ILES-256 vs ILES-512) and
$\approx 10\%$ (ILES-128 vs ILES-512), and the kurtosis by $\approx  10\%$ and 
$\approx 25\%$, respectively.

All eddy-viscosity and scale-similarity models follow this trend.
They evolve virtually identically to the ILES-128 ensemble.
Also the nonlinear model based on grid-scale quantities (NL-GS) does not have
a measurable impact on the results.
However, the explicitly filtered nonlinear models (NL-F3 and NL-F5) clearly
improve over the ensemble without model.
Their evolution is consistent with the higher resolution (ILES-256) results.
Moreover, the differences between using a three-point stencil and a five-point
stencil are negligible indicating a converged result.

\begin{figure}[htbp]
	\begin{center}
		\includegraphics{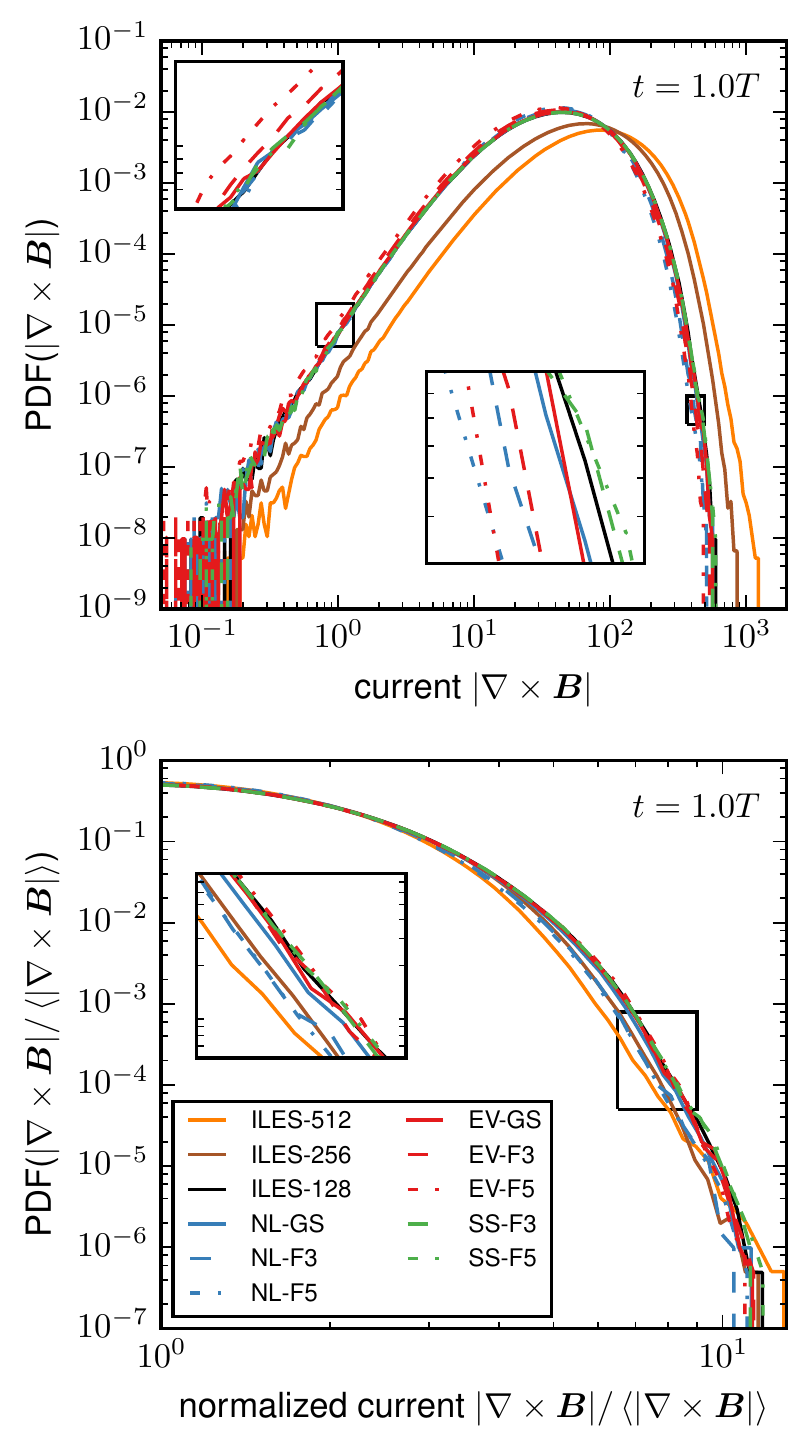}
	\end{center}
	\caption{Probability density function of the stacked 
		current density magnitude,
		i.e. each bin contains the 
		sum over all 19 realizations, at $t=1T$.
        The insets are magnifications of the indicated
        regions. The top panel shows the original raw data.
	The bottom panel details the right tail. 
	It has been normalized by the respective mean value of 
	each configuration in 
	order to highlight resolution independent features.
	}
	\label{fig:PDFJ}
\end{figure}

To further illustrate this we show the instantaneous probability density 
function (PDF) of the current magnitude at $t=1T$ in Fig.~\ref{fig:PDFJ}.
The top panel illustrates the raw PDFs.
The ensembles at different resolutions are clearly identified by 
an overall shift on the x-axis.
This corresponds to the decrease of the mean with resolution, as described 
in section~\ref{sec:resMeans}.
The differences in the higher order statistics 
of the different configurations are already hinted at 
in the insets.  
A pure shift would equally affect the left and right hand side
tails.
This is observed in the PDFs of EV-F3 and EV-F5, which are both shifted
to the left in comparison to the implicit LES configuration.
In contrast to this, the PDFs of NL-F3 and NL-F5 only exhibit a shift
in the right tail and coincide with the ILES-128 in the left tail indicating
a change in the shape of the PDF.

This difference is evident in the PDFs of the normalized current in 
the bottom panel of Fig.~\ref{fig:PDFJ}, 
where the individual PDFs have been normalized by 
the respective mean values.
The three features previously identified in the temporal evolution of the 
skewness and kurtosis, i.e. the resolution differences (ILES-128 vs ILES-256 vs 
ILES-512), the insensitivity of the EV and SS models, and the improvement by the
nonlinear model, are clearly present.
In fact, the ensemble distributions of the nonlinear model (NL-F3 and NL-F5)
at a resolution of $128^3$ match the distribution of the implicit LES 
at a resolution of $256^3$ demonstrating a clear enhancement.
Finally, we emphasize again that the results of the current density
presented in this subsection are qualitative identical to the ones
obtained for the vorticity and dilatation magnitude, i.e.
the explicitly filtered nonlinear models match the higher-resolution
ILES.

\subsection{Structure functions}
In order to gain further insight into the flow we analyze the
structure functions \citep[e.g.][]{biskamp}.
In particular, we look at the longitudinal velocity structure 
functions of order $p$
\begin{equation}
	S^p_\parallel(l) = \left < \abs{
	(\v{u}(\v{x} + \v{l}) - \v{u}(\v{x})) \cdot \v{l}/l
	}^p
	\right > 
\end{equation}
which are given by the moments of the velocity increments along
the direction of separation $\v{l}$ assuming homogeneity and isotropy.
Structure functions are related to the correlation functions and the 
energy spectrum. 
Moreover, they exhibit scaling behavior in the inertial range 
$S^p(l) \propto l^{\zeta_p}$ so that scaling exponents $\zeta_p$ 
can be determined.
\begin{figure}[htbp]
	\begin{center}
		\includegraphics{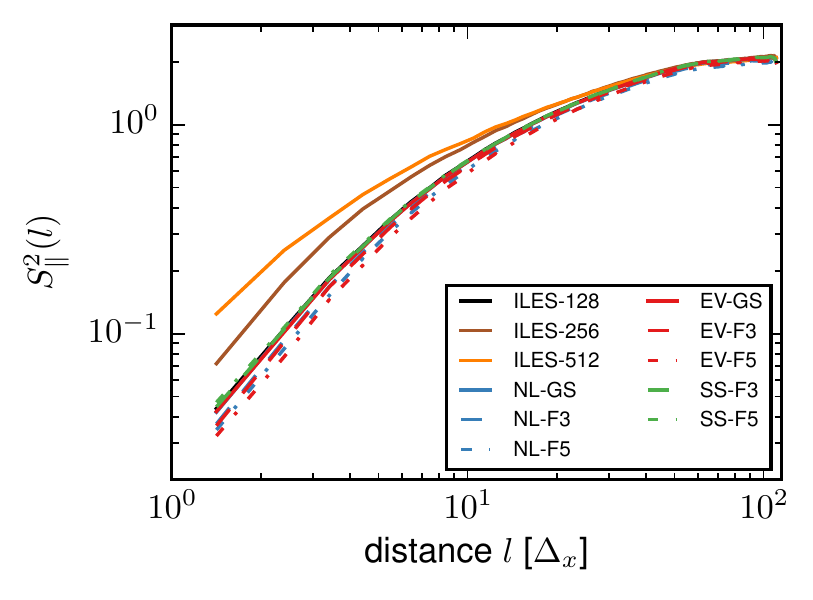}
		\caption{Second order longitudinal velocity
			structure function of one arbitrary 
			realization at $t=1T$.
	\label{fig:SFillusRaw}}
	\end{center}
\end{figure}
\begin{figure}[htbp]
	\begin{center}
			\includegraphics{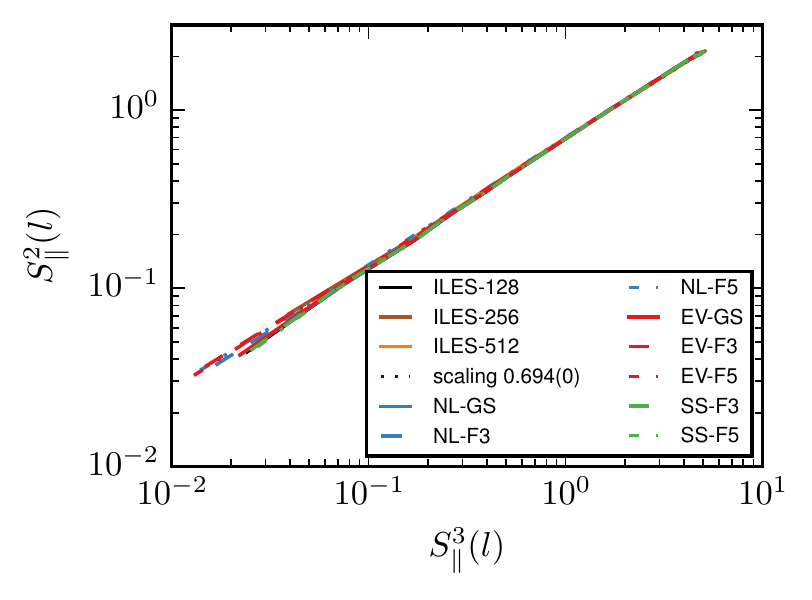}
	\end{center}
		\caption{Second order versus third order 
			longitudinal velocity
			structure function
		illustrating extended self-similarity of 
		the same realization as in Fig.~\ref{fig:SFillusRaw}.
		The best power-law fit (blue, dotted $\dotsb$) 
		to the ILES-512 simulation has an index of $0.694(0)$.
		\label{fig:SFillusESS}}
\end{figure}
\begin{figure}[htbp]
	\begin{center}
			\includegraphics{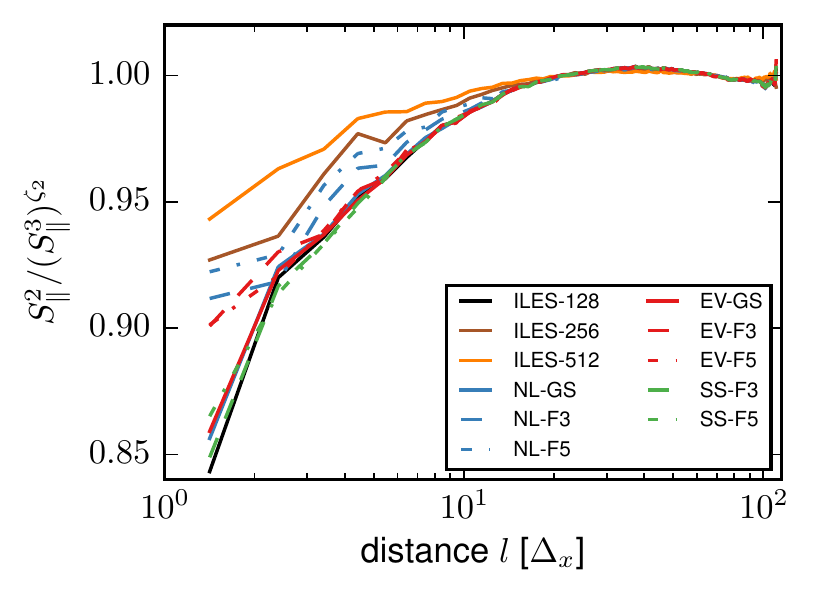}
	\end{center}
		\caption{Second order 
			longitudinal velocity
			structure functions normalized to the 
		third order structure function scaled by the best-fit 
	exponent. 
	The lines indicate the median over all 19 realizations
	at $t = 1T$.
	Variations as measured by the interquartile range are $<9\%$.
	\label{fig:SFillusRel}}
\end{figure}
Figure~\ref{fig:SFillusRaw} illustrates the second order longitudinal
structure function $S^2_\parallel$ for all configurations
of one arbitrary realization after one turnover time of free decay.
All structure functions have been calculated based on
$10^{10}$ randomly chosen pair of points.
The convergence has been verified by comparing the results with the ones 
obtained by using
twice the amount of points for one particular snapshot.
Two important features can be observed.
First, the structure functions of all configurations, i.e.
independent of resolution and presence of an SGS model, collapse
(on top of each other) on scales $\gtrsim 30 \Delta_x$.
On smaller scales, the differences with respect to resolution are 
more pronounced.
This is expected since the increasing numerical dissipation with decreasing 
resolution leads to a
decrease of variations in the velocity field on the small scales.
Again, the grid-filtered EV and NL LES, and the scale-similarity runs are
indistinguishable from the ILES run.
The increased dissipation of the EV and NL model already observed in the 
spectra and mean quantities is also visible here in the slightly reduced
variations on the smallest scales.
Second, no clear power-law range can be identified in any of the 
configurations which can be attributed to the limited resolution, which
for these simulations indicates a too low Reynolds number.

For this reason, we make use of the concept of extended self-similarity (ESS)
stating that the scaling behavior with 
corresponding scaling exponents can be recovered by relating structure functions
to each other.
While originally discovered in hydrodynamics \citep{PhysRevE.48.R29}, this
concept works remarkably well in MHD, too.
For example, in Fig.~\ref{fig:SFillusESS} we plot $S^2_{\parallel}$ versus 
$S^3_{\parallel}$ (of the same snapshot as in Fig.~\ref{fig:SFillusRaw}).
A power-law behavior for all configurations is clearly visible.
Moreover, the scaling exponents in this representations are by construction 
identical to the original ones.
Thus, we determine the individual exponents in this representation
by nonlinear least-square 
minimization using the \textsc{lmfit} package \citep{newville_2014_11813}. 
With these exponents, we now continue our analysis in two directions:
reevaluation of the structure functions versus separation distance, and
scaling behavior with increasing order $p$.

Figure~\ref{fig:SFillusRel} shows the median (over all 19 realization)
second order structure function of all configurations compensated by 
the corresponding third-order structure function and scaling exponent,
i.e. $S^2_{\parallel}/(S^3_\parallel)^{\zeta_2}$, versus distance $l$.
The plot illustrates where and to what extent the power-law scaling is found
in the non-normalized data.
Approximate power-law scaling is observed for all configurations on
scales $\gtrsim 20\Delta_x$.
Below $20\Delta_x$ the individual configurations start to deviate from
ideal scaling.
The deviations grow towards smaller scales for all configurations,
however, to different degrees for different configurations.
As expected, the highest resolution runs (ILES-512) exhibit the least deviation
(at most $6\%$ on the smallest scale), followed by the intermediate resolution
runs (ILES-256) with $\approx 7\%$ on the smallest scale.
At the lowest resolution, the no-model (ILES-128), grid-scale filtered
SGS (EV-GS and NL-GS) and scale-similarity (SS-F3 and SS-F5) runs show the
strongest deviation, $\approx 15\%$.
The explicitly filtered eddy-viscosity (EV-F3 and EV-F5) 
and nonlinear models (NL-F3 and NL-F5) display an improved behavior over
the other low resolution runs.
While the two EV models deviate by $10\%$, the nonlinear models deviate 
$8$-$9\%$, reaching almost the performance of the intermediate resolution runs.
Qualitatively, the same behavior observed for the structure functions of order 
$p=2$ is also observed for the structure functions of order $p=1$ and higher 
orders.

\begin{figure*}[htbp]
	\begin{center}
		\includegraphics{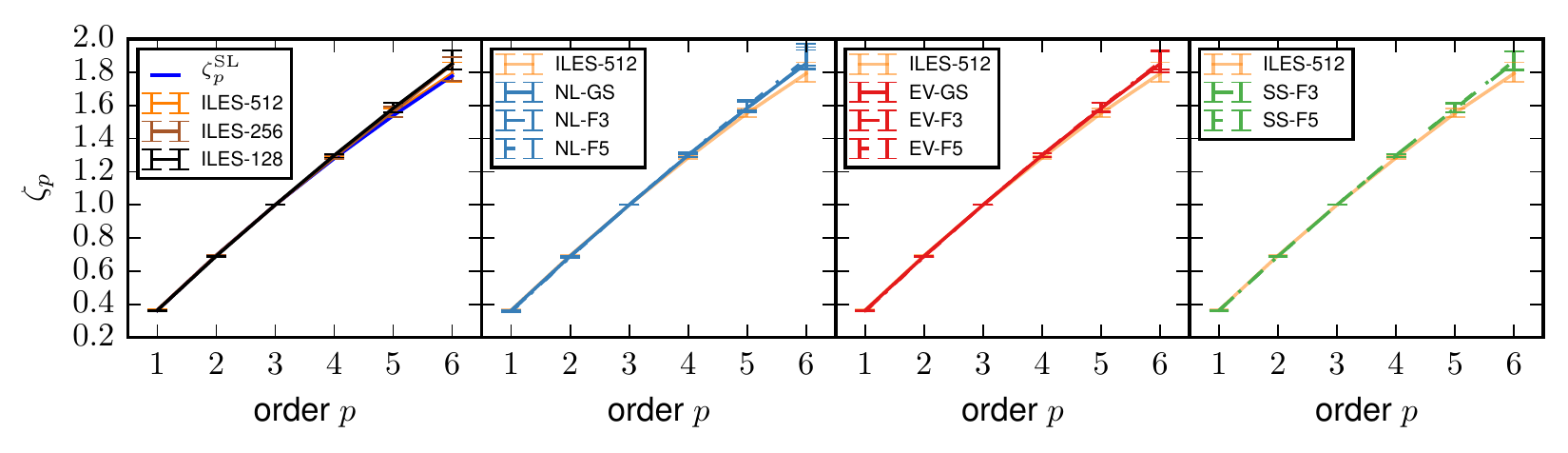}
	\end{center}
	\caption{Scaling exponents $\zeta_p$ of the longitudinal velocity structure 
		functions of order $p$ at $t=1T$.
	The lines indicate the median over all 19 realization and the errorbars 
	illustrate the interquartile range.
	For reference, the theoretical scaling $\zeta_p^\mathrm{SL}$ 
	derived by She and Leveque \citep{PhysRevLett.72.336} is shown in the
	left panel, and the reference simulation (ILES-512) is drawn in all panels.}
	\label{fig:SFscaling}
\end{figure*}
Finally, we analyze how the scaling exponents evolve with order $p$ depending
on resolution and presence of an SGS model.
Figure~\ref{fig:SFscaling} illustrates the median coefficients $\zeta_o$ over all 19 
realizations up to order $p=6$.
The coefficients have been determined based on nonlinear least-square fitting and 
employing extended self-similarity.
For higher orders the ESS does not provide robust exponents any more.
The left panel of Fig.~\ref{fig:SFscaling} shows the influence of decreasing 
resolution on the exponents.
While the exponents up to $p=4$ are virtually identical, a lower resolution 
(ILES-128 and ILES-256) leads to a slight overestimation of $\zeta_5$ ($2\%$)
and $\zeta_6$ ($3\%$) in comparison to ILES-512.
For reference, we also plot the exponents as derived by 
She and Leveque \citep{PhysRevLett.72.336} under the assumption that the most 
singular dissipative structures are filaments
\begin{equation}
	\zeta_p = \frac{p}{9} + 2 \bra{1 - \bra{\frac{2}{3}}^{p/3}} \;.
\end{equation}
The reference run ILES-512 fits the prediction remarkably well with a deviation of 
only $8\permil$ at the highest order $p=6$.
In general, the different SGS models and filtering procedures
do not have a measurable influence on the scaling behavior.
All behave like the ILES-128 yielding slightly overestimated exponents at high
order, which can be attributed entirely to the low resolution.

\subsection{Computational efficiency}
Finally, we compare the additional computational costs incurred
with the calculation of the different SGS models and 
the filtering.
Given that all LES were conducted at a resolution of $128^3$ grid
points, we compare the relative overhead over the no-model
ensemble at the same resolution (ILES-128).
Table~\ref{tab:CompEff} lists the mean ratios of the time per individual cycle
at the fluid level, i.e. other factors such as inter-process communication are
not included.
\begin{table}
	\caption{
Computational efficiency of the different SGS models relative to
the no-model run at identical resolution (ILES-128).
For reference, the efficiency of the no-model run at the next higher
resolution (ILES-256) is also shown.
The numbers represent the mean values (at the fluid level) 
over three test runs of 300 cycles each.
Each run used the same executable on a single machine employing
8 MPI-processes (no threading). 
This corresponds to a $64^3$ grid per process, as 
suggested by the \textsc{Enzo} documentation.
}
\label{tab:CompEff}
\begin{ruledtabular}
\begin{tabular}{lccc}
& GS & F3 & F5 \\
\noalign{\smallskip}\hline\noalign{\smallskip}
   NL  & 1.175(5) 	 & 1.468(5) 	& 2.411(7) 	\\
   EV  & 1.133(5) 	 & 1.431(6) 	& 2.368(8) 	\\
   SS  &		 & 1.949(8) 	& 4.804(16) 	\\
   ILES-256  & 9.242(28)\footnote{
Please note that this number only includes the time per cycle.
The total computational costs are increased by another factor of 
$\gtrsim2$ due to the decreased timestep at higher resolution.
   }& & \\
\end{tabular}
\end{ruledtabular}
\end{table}
The time per cycle increases for all SGS models when compared to the ILES-128 as expected.
Furthermore, two general trends are visible.

First, the computational costs increase with increasing filter width.
For example, the nonlinear model with grid-scale quantities increases the
time per cycle by a factor of $\approx1.18$ (NL-GS).
Explicit filtering introduces additional computations and is thus even more expensive,
i.e. a factor of $\approx1.47$ for NL-F3 and of $\approx2.41$ for NL-F5, respectively.
The unproportional increase in computational costs between F3 and  
F5 is easily explained by the unfavorable memory access in the filtering procedure.
For F5 the filter is build upon a stencil involving $5^3=125$ points resulting
in many accesses to non-contiguous memory and thus cache-misses.

Second, the eddy-viscosity and nonlinear model introduce a similar overhead,
with EV being a few percent cheaper than NL, while
the scale-similarity model is about twice as expensive as the other two models.
The latter is attributed to the additional (comparatively expensive, explicit) 
filter operations.
NL and EV only require filter operations on the 7 primary quantities
($\flt{\rho}, \fav{\v{u}}$ and $\flt{\v{B}}$).
The scale-similarity models also needs all filtered mixed quantities
($\widebreve{\fav{u_i}\fav{u_j}}$,  $\widebreve{\flt{B_i}\flt{B_j}}$, and
$\widebreve{\fav{\v{u}}\times\flt{\v{B}}}$) which involves 15 additional filter
operations in total.

Finally, we also tested how the computational costs increase for a 
no-model run at the next higher resolution (ILES-256).
At the level of a single fluid cycle the time increases by factor of
$\approx9$.
However, this does not yet take into account that the timestep is also 
reduced by a factor of $\gtrsim 2$ at $256^3$ versus $128^3$.
Thus, the total time required to reach a certain state in the simulation
is effectively increased by a factor of $\gtrsim 18$.

\section{Discussion}
\label{sec:discussion}
One of the most striking results from the analysis in the last section is
that models calculated from quantities at the grid-scale, i.e. EV-GS and
NL-GS, and scale-similarity models (SS-F3 and SS-F5) have no measurable
impact on the statistics of the flow.
The results for the grid-scale based models are in agreement with
findings for finite-difference schemes \citep{GHOSAL1996187,Chow2003366} and
for shock-capturing methods \citep{Garnier1999273}, i.e. numerics dominate
over (eddy-viscosity type) SGS models when no explicit filtering is applied.
However this does not explain the results for the scale-similarity closures, 
which employ an explicit filter.
A possible explanation for the absence of any observable effect 
(apart from a very short transient
behavior, e.g. in the kurtosis and skewness of the current at $t<0.1T$ as visible
in Fig.~\ref{fig:EvolHigherOrder}) is that the filter 
separation is still too small.
Physically, this translates to the statement that the modeling assumption 
of self-similar turbulence 
is not fulfilled on these numerically strongly damped scales.

Another observation concerns the convergence with filter width for
explicitly filtered eddy-viscosity (EV-F3 and EV-F5) and nonlinear
(NL-F3 NL-F5) models.
The energy spectra and mean quantities exhibit a small dependency
on the filter width indicating increased dissipative behavior 
with larger $\Delta$.
However, this is secondary from a practical point of view
because a smaller explicit filter is desirable for two reasons.
First, higher order statistics, e.g. skewness, kurtosis and normalized 
structure functions, show approximately converged results for F3 and F5.
Thus, the improvements over the ILES-128 can already be achieved with the
smaller explicit filter and the nonlinear model (NL-F3) while 
possessing a smaller intrinsic dissipation and being computationally 
more efficient than the F5 counterpart.
Second, larger explicit filters in their current form are impractical 
for actual LES in any case.

Due to the unfavorable memory access in the filtering, the computational
cost grows exponentially with increasing filter width.
In addition, more and more ghost zones are required increasing the
costs even further.
Building multi-dimensional filters based on the simultaneous application of one
dimensional ones rather than the sequential application could be a potential
way out.
Even though they are found to be slightly less accurate \citep{FLD:FLD914}, 
their multi-dimensional stencil size is decreased dramatically, i.e.
from $N^3$ to $3(N-1) + 1$ supporting points in three dimensions
with $N$ being the number of points for the 1-d filter.
Alternatively, the filtering could be realized in spectral space.
While the process of filtering itself is then reduced to a simple local
multiplication, additional complexity independent of the filter width,
is introduced by the transformations between real and spectral space.

Our current filtering framework could also be further optimized 
to reduce the computational overhead of the filtering.
For example, cache misses would be partly avoided by using fixed,
compiled-in stencils rather than dynamic ones in each cycle.
Independently of this, in practice the estimated SGS modeling overhead in a 
simulation is rather conservative. 
The total wallclock time always depends on additional  factors than
the time spent at the pure fluid level, most notably inter-process
communication.
Moreover, additional physics present in the simulation such as gravity or chemistry
can reduce the relative overhead introduced by an SGS model even further.
 
All models and filters lead to stable simulations and we did not employ any
explicitly regularization.
While this comes as no surprise for the eddy-viscosity type models, which are 
only capable of transferring energy down-scale, other groups, e.g. 
\citep{:/content/aip/journal/pof2/13/5/10.1063/1.1360192,Mueller2002},
typically find that regularization is required for scale-similarity and 
nonlinear type models.
These models also allow for up-scale energy transfer and are thus
capable of seeding numerical instabilities when this inverse transfer is
not controlled.
Most frequently, both type of models are therefore supplemented with an additional
eddy-viscosity type term which successfully stabilizes the simulations.
However, this only concerns (non shock-capturing) finite-volume,
finite difference or (pseudo-) spectral schemes.
In our case of a shock-capturing finite-volume scheme, the inherent numerical 
dissipation acts as an effective eddy-viscosity model (thus the term
implicit LES, see e.g. \citep{grinstein2007implicit}) and evidently
provides sufficient regularization for stable simulations.

It should be noted the present study only analyzes turbulence
with negligible cross-helicity.
In order to account for cross-helicity effects such as changing cascade 
dynamics both structural models are expected to be applicable as presented.
While the nonlinear model makes no assumptions on the nature of the 
flow \cite{Vlaykov2016a}, 
the scale-similarity model implicitly accounts for changing 
dynamics \cite{Chernyshov2012}.
The functional eddy-viscosity model misses this feature.
For this reason, extensions have been proposed to explicitly
capture unresolved cross-helicity effects \cite{Yokoi2013}.
Similarly, all models presented concern fully collisional plasmas.
Kinetic effects from low collisionality plasmas such as
anisotropic thermal conduction even in the presence of weak mean fields 
or heating and nonthermal particle acceleration from magnetic reconnection
are not captured by the presented models \cite{Miesch2015}.

Finally, the resolution of the LES ($128^3$) in this work prohibits the study
of more detailed physical effects in MHD turbulence itself \cite{Zhou2011}
and the corresponding properties of the SGS models.
For example, in order to analyze the locality of interactions \cite{Zhou20101}
a much higher resolution is required so that the effects of 
numerical dissipation can be clearly distinguished from physical ones.
In addition, higher resolution LES would also allow to study the properties
of the SGS models with respect to how they affect the straining (i.e. the
distortion of small scale vortices by large scale motions) and sweeping 
(i.e. advection of small scale vortices 
by the large scale flow with negligible distortions) 
in MHD turbulence \cite{Zhou2004}.
However, we expect that with increasing resolution the importance of a
pure SGS turbulence model decreases as more and more features are naturally 
resolved.
\section{Conclusions and outlook}
\label{sec:conclusions}
In this paper we analyzed the free decay of homogeneous, isotropic,
supersonic MHD turbulence with different SGS models and without
explicit model on various grid resolutions.
SGS models are typically introduced to LES in order to incorporate 
unaccounted effects from below the grid-scale, and to improve
the quality of the simulation at lower computational cost.
We measured the quality of the SGS models by their capability
to reproduce the results of a reference simulation at higher resolution.
The reference quantities included the energy spectra, the evolution 
of different statistical moments of 
the kinetic and magnetic energies, the vorticity, the current density and 
the dilatation magnitudes.
In total, we compared three SGS models: eddy-viscosity, scale-similarity and 
nonlinear. 
Additionally, we evaluated the influence of using
implicit-, grid-filtered quantities versus
explicitly filtered quantities to calculate the model terms.
We analyzed an ensemble of 19 different initial conditions
for each configuration as temporary, transient fluctuations 
can easily dominate individual simulations.

We find that the simulations employing a grid-filtered eddy-viscosity (EV-GS)
or nonlinear (NL-GS) model, or a scale-similarity model with the tested explicit 
filters (SS-F3 and SS-F5) produce results that are indistinguishable from an implicit
LES, i.e. without an explicit model, at the same resolution.
Moreover, we find that the eddy-viscosity and nonlinear models
with the two tested explicit filter widths, i.e. with filter widths
of $2.71\Delta_x$ (EV-F3 and NL-F3) and $4.75\Delta_x$ (EV-F5 and NL-F5), 
introduce little additional
dissipation on the smallest represented scales e.g. in the energy
spectra or the evolution of the mean quantities.
Finally, the nonlinear model (NL-F3 and NL-F5) improves higher order 
statistics of
small-scale dependent quantities, such as the kurtosis and skewness
of the current density, dilatation and vorticity.
For these quantities, the results of an ILES at doubled resolution (in
each dimension) can be achieved while introducing only a small computational 
overhead --- less than factor of $1.5$ (versus $\approx16$ for the higher-resolution
ILES). 
This similarly applies to the normalized structure functions and is independent
of the explicit filter width.

Based on these results we conclude that an explicit filter is required in order
to obtain a measurable impact of an SGS model for
shock-capturing finite-volume schemes of second order.
In how far this conclusion holds for schemes of higher order and more dynamic 
versions of the SGS models, e.g. with dynamic coefficients, is yet to be seen 
and subject to future work.
Furthermore, additional dissipation for the explicitly filtered models is
not required as numerical dissipation proves to be sufficient (if not too high).
Thus, the introduction of an eddy-viscosity model in these schemes is unnecessary.
However, as the nonlinear model improves higher order statistics, it would be
desirable to remove the unnecessary dissipation to improve the lower-order
statistics as well.
This kind of regularization is also subject of future work.
Finally, the nonlinear model in its current version can readily be used with
a small explicit filter in situation where higher order statistics are important
at little extra cost.
The associated code will be made publicly available together with the publication.

\begin{acknowledgments}
  PG acknowledges financial support by the
  \textit{International Max Planck Research School for Solar System 
  Science at the University of G\"ottingen}.
  DV acknowledge research funding by the
  Max-Planck-Institut f\"ur Dynamik und Selbstorganisation.  
  DRGS thanks for funding through Fondecyt regular (project code 1161247),
  through the ''Concurso Proyectos Internacionales de Investigaci\'on,
  Convocatoria 2015'' (project code PII20150171), and
  from the Chilean BASAL Centro de Excelencia en Astrof\'isica y Tecnolog\'as 
  Afines (CATA) grant PFB-06/2007.
  The \Enzo simulations were performed and analyzed with the HLRN-III facilities of
  the \textit{North-German Supercomputing Alliance} under grant \textit{nip00037}.
\end{acknowledgments}


\begin{thebibliography}{47}%
\makeatletter
\providecommand \@ifxundefined [1]{%
 \@ifx{#1\undefined}
}%
\providecommand \@ifnum [1]{%
 \ifnum #1\expandafter \@firstoftwo
 \else \expandafter \@secondoftwo
 \fi
}%
\providecommand \@ifx [1]{%
 \ifx #1\expandafter \@firstoftwo
 \else \expandafter \@secondoftwo
 \fi
}%
\providecommand \natexlab [1]{#1}%
\providecommand \enquote  [1]{``#1''}%
\providecommand \bibnamefont  [1]{#1}%
\providecommand \bibfnamefont [1]{#1}%
\providecommand \citenamefont [1]{#1}%
\providecommand \href@noop [0]{\@secondoftwo}%
\providecommand \href [0]{\begingroup \@sanitize@url \@href}%
\providecommand \@href[1]{\@@startlink{#1}\@@href}%
\providecommand \@@href[1]{\endgroup#1\@@endlink}%
\providecommand \@sanitize@url [0]{\catcode `\\12\catcode `\$12\catcode
  `\&12\catcode `\#12\catcode `\^12\catcode `\_12\catcode `\%12\relax}%
\providecommand \@@startlink[1]{}%
\providecommand \@@endlink[0]{}%
\providecommand \url  [0]{\begingroup\@sanitize@url \@url }%
\providecommand \@url [1]{\endgroup\@href {#1}{\urlprefix }}%
\providecommand \urlprefix  [0]{URL }%
\providecommand \Eprint [0]{\href }%
\providecommand \doibase [0]{http://dx.doi.org/}%
\providecommand \selectlanguage [0]{\@gobble}%
\providecommand \bibinfo  [0]{\@secondoftwo}%
\providecommand \bibfield  [0]{\@secondoftwo}%
\providecommand \translation [1]{[#1]}%
\providecommand \BibitemOpen [0]{}%
\providecommand \bibitemStop [0]{}%
\providecommand \bibitemNoStop [0]{.\EOS\space}%
\providecommand \EOS [0]{\spacefactor3000\relax}%
\providecommand \BibitemShut  [1]{\csname bibitem#1\endcsname}%
\let\auto@bib@innerbib\@empty
\bibitem [{\citenamefont {Balbus}\ and\ \citenamefont
  {Hawley}(1998)}]{RevModPhys.70.1}%
  \BibitemOpen
  \bibfield  {author} {\bibinfo {author} {\bibfnamefont {S.~A.}\ \bibnamefont
  {Balbus}}\ and\ \bibinfo {author} {\bibfnamefont {J.~F.}\ \bibnamefont
  {Hawley}},\ }\href {\doibase 10.1103/RevModPhys.70.1} {\bibfield  {journal}
  {\bibinfo  {journal} {Rev. Mod. Phys.}\ }\textbf {\bibinfo {volume} {70}},\
  \bibinfo {pages} {1} (\bibinfo {year} {1998})}\BibitemShut {NoStop}%
\bibitem [{\citenamefont {{Goldstein}}\ \emph {et~al.}(1995)\citenamefont
  {{Goldstein}}, \citenamefont {{Roberts}},\ and\ \citenamefont
  {{Matthaeus}}}]{1995ARA&A..33..283G}%
  \BibitemOpen
  \bibfield  {author} {\bibinfo {author} {\bibfnamefont {M.~L.}\ \bibnamefont
  {{Goldstein}}}, \bibinfo {author} {\bibfnamefont {D.~A.}\ \bibnamefont
  {{Roberts}}}, \ and\ \bibinfo {author} {\bibfnamefont {W.~H.}\ \bibnamefont
  {{Matthaeus}}},\ }\href {\doibase 10.1146/annurev.aa.33.090195.001435}
  {\bibfield  {journal} {\bibinfo  {journal} {\araa}\ }\textbf {\bibinfo
  {volume} {33}},\ \bibinfo {pages} {283} (\bibinfo {year} {1995})}\BibitemShut
  {NoStop}%
\bibitem [{\citenamefont {{Rodenbeck, Kai}}\ and\ \citenamefont {{Schleicher,
  Dominik R. G.}}(2016)}]{Rodenbeck2016}%
  \BibitemOpen
  \bibfield  {author} {\bibinfo {author} {\bibnamefont {{Rodenbeck, Kai}}}\
  and\ \bibinfo {author} {\bibnamefont {{Schleicher, Dominik R. G.}}},\ }\href
  {\doibase 10.1051/0004-6361/201527393} {\bibfield  {journal} {\bibinfo
  {journal} {A\&A}\ }\textbf {\bibinfo {volume} {593}},\ \bibinfo {pages} {A89}
  (\bibinfo {year} {2016})}\BibitemShut {NoStop}%
\bibitem [{\citenamefont {Oishi}\ \emph {et~al.}(2015)\citenamefont {Oishi},
  \citenamefont {Low}, \citenamefont {Collins},\ and\ \citenamefont
  {Tamura}}]{2015ApJ...806L..12O}%
  \BibitemOpen
  \bibfield  {author} {\bibinfo {author} {\bibfnamefont {J.~S.}\ \bibnamefont
  {Oishi}}, \bibinfo {author} {\bibfnamefont {M.-M.~M.}\ \bibnamefont {Low}},
  \bibinfo {author} {\bibfnamefont {D.~C.}\ \bibnamefont {Collins}}, \ and\
  \bibinfo {author} {\bibfnamefont {M.}~\bibnamefont {Tamura}},\ }\href
  {http://stacks.iop.org/2041-8205/806/i=1/a=L12} {\bibfield  {journal}
  {\bibinfo  {journal} {The Astrophysical Journal Letters}\ }\textbf {\bibinfo
  {volume} {806}},\ \bibinfo {pages} {L12} (\bibinfo {year}
  {2015})}\BibitemShut {NoStop}%
\bibitem [{\citenamefont {Schober}\ \emph {et~al.}(2015)\citenamefont
  {Schober}, \citenamefont {Schleicher}, \citenamefont {Federrath},
  \citenamefont {Bovino},\ and\ \citenamefont {Klessen}}]{PhysRevE.92.023010}%
  \BibitemOpen
  \bibfield  {author} {\bibinfo {author} {\bibfnamefont {J.}~\bibnamefont
  {Schober}}, \bibinfo {author} {\bibfnamefont {D.~R.~G.}\ \bibnamefont
  {Schleicher}}, \bibinfo {author} {\bibfnamefont {C.}~\bibnamefont
  {Federrath}}, \bibinfo {author} {\bibfnamefont {S.}~\bibnamefont {Bovino}}, \
  and\ \bibinfo {author} {\bibfnamefont {R.~S.}\ \bibnamefont {Klessen}},\
  }\href {\doibase 10.1103/PhysRevE.92.023010} {\bibfield  {journal} {\bibinfo
  {journal} {Phys. Rev. E}\ }\textbf {\bibinfo {volume} {92}},\ \bibinfo
  {pages} {023010} (\bibinfo {year} {2015})}\BibitemShut {NoStop}%
\bibitem [{\citenamefont {{Cooper}}\ \emph {et~al.}(2014)\citenamefont
  {{Cooper}}, \citenamefont {{Wallace}}, \citenamefont {{Brookhart}},
  \citenamefont {{Clark}}, \citenamefont {{Collins}}, \citenamefont {{Ding}},
  \citenamefont {{Flanagan}}, \citenamefont {{Khalzov}}, \citenamefont {{Li}},
  \citenamefont {{Milhone}}, \citenamefont {{Nornberg}}, \citenamefont
  {{Nonn}}, \citenamefont {{Weisberg}}, \citenamefont {{Whyte}}, \citenamefont
  {{Zweibel}},\ and\ \citenamefont {{Forest}}}]{2014PhPl...21a3505C}%
  \BibitemOpen
  \bibfield  {author} {\bibinfo {author} {\bibfnamefont {C.~M.}\ \bibnamefont
  {{Cooper}}}, \bibinfo {author} {\bibfnamefont {J.}~\bibnamefont {{Wallace}}},
  \bibinfo {author} {\bibfnamefont {M.}~\bibnamefont {{Brookhart}}}, \bibinfo
  {author} {\bibfnamefont {M.}~\bibnamefont {{Clark}}}, \bibinfo {author}
  {\bibfnamefont {C.}~\bibnamefont {{Collins}}}, \bibinfo {author}
  {\bibfnamefont {W.~X.}\ \bibnamefont {{Ding}}}, \bibinfo {author}
  {\bibfnamefont {K.}~\bibnamefont {{Flanagan}}}, \bibinfo {author}
  {\bibfnamefont {I.}~\bibnamefont {{Khalzov}}}, \bibinfo {author}
  {\bibfnamefont {Y.}~\bibnamefont {{Li}}}, \bibinfo {author} {\bibfnamefont
  {J.}~\bibnamefont {{Milhone}}}, \bibinfo {author} {\bibfnamefont
  {M.}~\bibnamefont {{Nornberg}}}, \bibinfo {author} {\bibfnamefont
  {P.}~\bibnamefont {{Nonn}}}, \bibinfo {author} {\bibfnamefont
  {D.}~\bibnamefont {{Weisberg}}}, \bibinfo {author} {\bibfnamefont {D.~G.}\
  \bibnamefont {{Whyte}}}, \bibinfo {author} {\bibfnamefont {E.}~\bibnamefont
  {{Zweibel}}}, \ and\ \bibinfo {author} {\bibfnamefont {C.~B.}\ \bibnamefont
  {{Forest}}},\ }\href {\doibase 10.1063/1.4861609} {\bibfield  {journal}
  {\bibinfo  {journal} {Physics of Plasmas}\ }\textbf {\bibinfo {volume}
  {21}},\ \bibinfo {eid} {013505} (\bibinfo {year} {2014})}\BibitemShut
  {NoStop}%
\bibitem [{\citenamefont {Tzeferacos}\ \emph {et~al.}(2015)\citenamefont
  {Tzeferacos}, \citenamefont {Fatenejad}, \citenamefont {Flocke},
  \citenamefont {Graziani}, \citenamefont {Gregori}, \citenamefont {Lamb},
  \citenamefont {Lee}, \citenamefont {Meinecke}, \citenamefont {Scopatz},\ and\
  \citenamefont {Weide}}]{Tzeferacos201524}%
  \BibitemOpen
  \bibfield  {author} {\bibinfo {author} {\bibfnamefont {P.}~\bibnamefont
  {Tzeferacos}}, \bibinfo {author} {\bibfnamefont {M.}~\bibnamefont
  {Fatenejad}}, \bibinfo {author} {\bibfnamefont {N.}~\bibnamefont {Flocke}},
  \bibinfo {author} {\bibfnamefont {C.}~\bibnamefont {Graziani}}, \bibinfo
  {author} {\bibfnamefont {G.}~\bibnamefont {Gregori}}, \bibinfo {author}
  {\bibfnamefont {D.}~\bibnamefont {Lamb}}, \bibinfo {author} {\bibfnamefont
  {D.}~\bibnamefont {Lee}}, \bibinfo {author} {\bibfnamefont {J.}~\bibnamefont
  {Meinecke}}, \bibinfo {author} {\bibfnamefont {A.}~\bibnamefont {Scopatz}}, \
  and\ \bibinfo {author} {\bibfnamefont {K.}~\bibnamefont {Weide}},\ }\href
  {\doibase 10.1016/j.hedp.2014.11.003} {\bibfield  {journal} {\bibinfo
  {journal} {High Energy Density Physics}\ }\textbf {\bibinfo {volume} {17,
  Part A}},\ \bibinfo {pages} {24 } (\bibinfo {year} {2015})},\ \bibinfo {note}
  {special Issue: 10th International Conference on High Energy Density
  Laboratory Astrophysics}\BibitemShut {NoStop}%
\bibitem [{\citenamefont {Kaneda}\ \emph {et~al.}(2003)\citenamefont {Kaneda},
  \citenamefont {Ishihara}, \citenamefont {Yokokawa}, \citenamefont {Itakura},\
  and\ \citenamefont {Uno}}]{Kaneda2003}%
  \BibitemOpen
  \bibfield  {author} {\bibinfo {author} {\bibfnamefont {Y.}~\bibnamefont
  {Kaneda}}, \bibinfo {author} {\bibfnamefont {T.}~\bibnamefont {Ishihara}},
  \bibinfo {author} {\bibfnamefont {M.}~\bibnamefont {Yokokawa}}, \bibinfo
  {author} {\bibfnamefont {K.}~\bibnamefont {Itakura}}, \ and\ \bibinfo
  {author} {\bibfnamefont {A.}~\bibnamefont {Uno}},\ }\href {\doibase
  10.1063/1.1539855} {\bibfield  {journal} {\bibinfo  {journal} {Physics of
  Fluids}\ }\textbf {\bibinfo {volume} {15}},\ \bibinfo {pages} {L21} (\bibinfo
  {year} {2003})},\ \Eprint
  {http://arxiv.org/abs/http://aip.scitation.org/doi/pdf/10.1063/1.1539855}
  {http://aip.scitation.org/doi/pdf/10.1063/1.1539855} \BibitemShut {NoStop}%
\bibitem [{\citenamefont {Yeung}\ \emph {et~al.}(2015)\citenamefont {Yeung},
  \citenamefont {Zhai},\ and\ \citenamefont {Sreenivasan}}]{Yeung13102015}%
  \BibitemOpen
  \bibfield  {author} {\bibinfo {author} {\bibfnamefont {P.~K.}\ \bibnamefont
  {Yeung}}, \bibinfo {author} {\bibfnamefont {X.~M.}\ \bibnamefont {Zhai}}, \
  and\ \bibinfo {author} {\bibfnamefont {K.~R.}\ \bibnamefont {Sreenivasan}},\
  }\href {\doibase 10.1073/pnas.1517368112} {\bibfield  {journal} {\bibinfo
  {journal} {Proceedings of the National Academy of Sciences}\ }\textbf
  {\bibinfo {volume} {112}},\ \bibinfo {pages} {12633} (\bibinfo {year}
  {2015})},\ \Eprint
  {http://arxiv.org/abs/http://www.pnas.org/content/112/41/12633.full.pdf}
  {http://www.pnas.org/content/112/41/12633.full.pdf} \BibitemShut {NoStop}%
\bibitem [{\citenamefont {{Federrath}}\ \emph {et~al.}(2016)\citenamefont
  {{Federrath}}, \citenamefont {{Klessen}}, \citenamefont {{Iapichino}},\ and\
  \citenamefont {{Hammer}}}]{Federrath2016}%
  \BibitemOpen
  \bibfield  {author} {\bibinfo {author} {\bibfnamefont {C.}~\bibnamefont
  {{Federrath}}}, \bibinfo {author} {\bibfnamefont {R.~S.}\ \bibnamefont
  {{Klessen}}}, \bibinfo {author} {\bibfnamefont {L.}~\bibnamefont
  {{Iapichino}}}, \ and\ \bibinfo {author} {\bibfnamefont {N.~J.}\ \bibnamefont
  {{Hammer}}},\ }\enquote {\bibinfo {title} {The world's largest turbulence
  simulations},}\ in\ \href {http://arxiv.org/abs/1607.00630} {\emph {\bibinfo
  {booktitle} {High Performance Computing in Science und Engineering –
  Garching/Munich 2016}}},\ \bibinfo {editor} {edited by\ \bibinfo {editor}
  {\bibfnamefont {S.}~\bibnamefont {Wagner}}, \bibinfo {editor} {\bibfnamefont
  {A.}~\bibnamefont {Bode}}, \bibinfo {editor} {\bibfnamefont {H.}~\bibnamefont
  {Brüchle}}, \ and\ \bibinfo {editor} {\bibfnamefont {M.}~\bibnamefont
  {Brehm}}}\ (\bibinfo  {publisher} {Bayerische Akademie der Wissenschaften},\
  \bibinfo {year} {2016})\ Chap.~\bibinfo {chapter} {1}, pp.\ \bibinfo {pages}
  {30--31}\BibitemShut {NoStop}%
\bibitem [{\citenamefont {Sagaut}(2006)}]{Sagaut2006}%
  \BibitemOpen
  \bibfield  {author} {\bibinfo {author} {\bibfnamefont {P.}~\bibnamefont
  {Sagaut}},\ }\href
  {http://www.springer.com/de/book/9783540263449?wt_mc=ThirdParty.SpringerLink.3.EPR653.About_eBook}
  {\emph {\bibinfo {title} {Large Eddy Simulation for Incompressible Flows: An
  Introduction}}},\ Scientific Computation\ (\bibinfo  {publisher} {Springer},\
  \bibinfo {year} {2006})\BibitemShut {NoStop}%
\bibitem [{\citenamefont {Schmidt}(2015)}]{lrca-2015-2}%
  \BibitemOpen
  \bibfield  {author} {\bibinfo {author} {\bibfnamefont {W.}~\bibnamefont
  {Schmidt}},\ }\href {\doibase 10.1007/lrca-2015-2} {\bibfield  {journal}
  {\bibinfo  {journal} {Living Reviews in Computational Astrophysics}\ }\textbf
  {\bibinfo {volume} {1}} (\bibinfo {year} {2015}),\
  10.1007/lrca-2015-2}\BibitemShut {NoStop}%
\bibitem [{\citenamefont {Vlaykov}\ \emph {et~al.}(2016)\citenamefont
  {Vlaykov}, \citenamefont {Grete}, \citenamefont {Schmidt},\ and\
  \citenamefont {Schleicher}}]{Vlaykov2016a}%
  \BibitemOpen
  \bibfield  {author} {\bibinfo {author} {\bibfnamefont {D.~G.}\ \bibnamefont
  {Vlaykov}}, \bibinfo {author} {\bibfnamefont {P.}~\bibnamefont {Grete}},
  \bibinfo {author} {\bibfnamefont {W.}~\bibnamefont {Schmidt}}, \ and\
  \bibinfo {author} {\bibfnamefont {D.~R.~G.}\ \bibnamefont {Schleicher}},\
  }\href {\doibase 10.1063/1.4954303} {\bibfield  {journal} {\bibinfo
  {journal} {Physics of Plasmas}\ }\textbf {\bibinfo {volume} {23}},\ \bibinfo
  {eid} {062316} (\bibinfo {year} {2016}),\ 10.1063/1.4954303}\BibitemShut
  {NoStop}%
\bibitem [{\citenamefont {{Favre}}(1983)}]{1983PhFl...26.2851F}%
  \BibitemOpen
  \bibfield  {author} {\bibinfo {author} {\bibfnamefont {A.}~\bibnamefont
  {{Favre}}},\ }\href {\doibase 10.1063/1.864049} {\bibfield  {journal}
  {\bibinfo  {journal} {Physics of Fluids}\ }\textbf {\bibinfo {volume} {26}},\
  \bibinfo {pages} {2851} (\bibinfo {year} {1983})}\BibitemShut {NoStop}%
\bibitem [{\citenamefont {{Garnier}}\ \emph {et~al.}(2009)\citenamefont
  {{Garnier}}, \citenamefont {{Adams}},\ and\ \citenamefont
  {{Sagaut}}}]{2009lesc.book.....G}%
  \BibitemOpen
  \bibfield  {author} {\bibinfo {author} {\bibfnamefont {E.}~\bibnamefont
  {{Garnier}}}, \bibinfo {author} {\bibfnamefont {N.}~\bibnamefont {{Adams}}},
  \ and\ \bibinfo {author} {\bibfnamefont {P.}~\bibnamefont {{Sagaut}}},\
  }\href {\doibase 10.1007/978-90-481-2819-8} {\emph {\bibinfo {title} {{Large
  Eddy Simulation for Compressible Flows}}}},\ Scientific Computation\
  (\bibinfo  {publisher} {Springer Netherlands},\ \bibinfo {year}
  {2009})\BibitemShut {NoStop}%
\bibitem [{\citenamefont {Chernyshov}\ \emph {et~al.}(2014)\citenamefont
  {Chernyshov}, \citenamefont {Karelsky},\ and\ \citenamefont
  {Petrosyan}}]{Chernyshov2014}%
  \BibitemOpen
  \bibfield  {author} {\bibinfo {author} {\bibfnamefont {A.~A.}\ \bibnamefont
  {Chernyshov}}, \bibinfo {author} {\bibfnamefont {K.~V.}\ \bibnamefont
  {Karelsky}}, \ and\ \bibinfo {author} {\bibfnamefont {A.~S.}\ \bibnamefont
  {Petrosyan}},\ }\href {\doibase 10.3367/UFNe.0184.201405a.0457} {\bibfield
  {journal} {\bibinfo  {journal} {Physics-Uspekhi}\ }\textbf {\bibinfo {volume}
  {57}},\ \bibinfo {pages} {421} (\bibinfo {year} {2014})}\BibitemShut
  {NoStop}%
\bibitem [{\citenamefont {Miesch}\ \emph {et~al.}(2015)\citenamefont {Miesch},
  \citenamefont {Matthaeus}, \citenamefont {Brandenburg}, \citenamefont
  {Petrosyan}, \citenamefont {Pouquet}, \citenamefont {Cambon}, \citenamefont
  {Jenko}, \citenamefont {Uzdensky}, \citenamefont {Stone}, \citenamefont
  {Tobias}, \citenamefont {Toomre},\ and\ \citenamefont {Velli}}]{Miesch2015}%
  \BibitemOpen
  \bibfield  {author} {\bibinfo {author} {\bibfnamefont {M.}~\bibnamefont
  {Miesch}}, \bibinfo {author} {\bibfnamefont {W.}~\bibnamefont {Matthaeus}},
  \bibinfo {author} {\bibfnamefont {A.}~\bibnamefont {Brandenburg}}, \bibinfo
  {author} {\bibfnamefont {A.}~\bibnamefont {Petrosyan}}, \bibinfo {author}
  {\bibfnamefont {A.}~\bibnamefont {Pouquet}}, \bibinfo {author} {\bibfnamefont
  {C.}~\bibnamefont {Cambon}}, \bibinfo {author} {\bibfnamefont
  {F.}~\bibnamefont {Jenko}}, \bibinfo {author} {\bibfnamefont
  {D.}~\bibnamefont {Uzdensky}}, \bibinfo {author} {\bibfnamefont
  {J.}~\bibnamefont {Stone}}, \bibinfo {author} {\bibfnamefont
  {S.}~\bibnamefont {Tobias}}, \bibinfo {author} {\bibfnamefont
  {J.}~\bibnamefont {Toomre}}, \ and\ \bibinfo {author} {\bibfnamefont
  {M.}~\bibnamefont {Velli}},\ }\href {\doibase 10.1007/s11214-015-0190-7}
  {\bibfield  {journal} {\bibinfo  {journal} {Space Science Reviews}\ ,\
  \bibinfo {pages} {1}} (\bibinfo {year} {2015})}\BibitemShut {NoStop}%
\bibitem [{\citenamefont {Miki}\ and\ \citenamefont {Menon}(2008)}]{Miki2008}%
  \BibitemOpen
  \bibfield  {author} {\bibinfo {author} {\bibfnamefont {K.}~\bibnamefont
  {Miki}}\ and\ \bibinfo {author} {\bibfnamefont {S.}~\bibnamefont {Menon}},\
  }\href {\doibase 10.1063/1.2947312} {\bibfield  {journal} {\bibinfo
  {journal} {Physics of Plasmas (1994-present)}\ }\textbf {\bibinfo {volume}
  {15}},\ \bibinfo {eid} {072306} (\bibinfo {year} {2008})}\BibitemShut
  {NoStop}%
\bibitem [{\citenamefont {{Theobald}}\ \emph {et~al.}(1994)\citenamefont
  {{Theobald}}, \citenamefont {{Fox}},\ and\ \citenamefont
  {{Sofia}}}]{1994PhPl....1.3016T}%
  \BibitemOpen
  \bibfield  {author} {\bibinfo {author} {\bibfnamefont {M.~L.}\ \bibnamefont
  {{Theobald}}}, \bibinfo {author} {\bibfnamefont {P.~A.}\ \bibnamefont
  {{Fox}}}, \ and\ \bibinfo {author} {\bibfnamefont {S.}~\bibnamefont
  {{Sofia}}},\ }\href {\doibase 10.1063/1.870542} {\bibfield  {journal}
  {\bibinfo  {journal} {Physics of Plasmas}\ }\textbf {\bibinfo {volume} {1}},\
  \bibinfo {pages} {3016} (\bibinfo {year} {1994})}\BibitemShut {NoStop}%
\bibitem [{\citenamefont {Müller}\ and\ \citenamefont
  {Carati}(2002)}]{Mueller2002}%
  \BibitemOpen
  \bibfield  {author} {\bibinfo {author} {\bibfnamefont {W.-C.}\ \bibnamefont
  {Müller}}\ and\ \bibinfo {author} {\bibfnamefont {D.}~\bibnamefont
  {Carati}},\ }\href {\doibase 10.1063/1.1448498} {\bibfield  {journal}
  {\bibinfo  {journal} {Physics of Plasmas (1994-present)}\ }\textbf {\bibinfo
  {volume} {9}},\ \bibinfo {pages} {824} (\bibinfo {year} {2002})}\BibitemShut
  {NoStop}%
\bibitem [{\citenamefont {Yokoi}(2013)}]{Yokoi2013}%
  \BibitemOpen
  \bibfield  {author} {\bibinfo {author} {\bibfnamefont {N.}~\bibnamefont
  {Yokoi}},\ }\href {\doibase 10.1080/03091929.2012.754022} {\bibfield
  {journal} {\bibinfo  {journal} {Geophysical \& Astrophysical Fluid Dynamics}\
  }\textbf {\bibinfo {volume} {107}},\ \bibinfo {pages} {114} (\bibinfo {year}
  {2013})}\BibitemShut {NoStop}%
\bibitem [{\citenamefont {Grinstein}\ \emph {et~al.}(2007)\citenamefont
  {Grinstein}, \citenamefont {Margolin},\ and\ \citenamefont
  {Rider}}]{grinstein2007implicit}%
  \BibitemOpen
  \bibfield  {author} {\bibinfo {author} {\bibfnamefont {F.}~\bibnamefont
  {Grinstein}}, \bibinfo {author} {\bibfnamefont {L.}~\bibnamefont {Margolin}},
  \ and\ \bibinfo {author} {\bibfnamefont {W.}~\bibnamefont {Rider}},\ }\href
  {https://books.google.de/books?id=Xk-eb9kPgXsC} {\emph {\bibinfo {title}
  {Implicit Large Eddy Simulation: Computing Turbulent Fluid Dynamics}}}\
  (\bibinfo  {publisher} {Cambridge University Press},\ \bibinfo {year}
  {2007})\BibitemShut {NoStop}%
\bibitem [{\citenamefont {Grete}\ \emph {et~al.}(2015)\citenamefont {Grete},
  \citenamefont {Vlaykov}, \citenamefont {Schmidt}, \citenamefont
  {Schleicher},\ and\ \citenamefont {Federrath}}]{1367-2630-17-2-023070}%
  \BibitemOpen
  \bibfield  {author} {\bibinfo {author} {\bibfnamefont {P.}~\bibnamefont
  {Grete}}, \bibinfo {author} {\bibfnamefont {D.~G.}\ \bibnamefont {Vlaykov}},
  \bibinfo {author} {\bibfnamefont {W.}~\bibnamefont {Schmidt}}, \bibinfo
  {author} {\bibfnamefont {D.~R.~G.}\ \bibnamefont {Schleicher}}, \ and\
  \bibinfo {author} {\bibfnamefont {C.}~\bibnamefont {Federrath}},\ }\href
  {\doibase 10.1088/1367-2630/17/2/023070} {\bibfield  {journal} {\bibinfo
  {journal} {New Journal of Physics}\ }\textbf {\bibinfo {volume} {17}},\
  \bibinfo {pages} {023070} (\bibinfo {year} {2015})}\BibitemShut {NoStop}%
\bibitem [{\citenamefont {Grete}\ \emph {et~al.}(2016)\citenamefont {Grete},
  \citenamefont {Vlaykov}, \citenamefont {Schmidt},\ and\ \citenamefont
  {Schleicher}}]{Grete2016a}%
  \BibitemOpen
  \bibfield  {author} {\bibinfo {author} {\bibfnamefont {P.}~\bibnamefont
  {Grete}}, \bibinfo {author} {\bibfnamefont {D.~G.}\ \bibnamefont {Vlaykov}},
  \bibinfo {author} {\bibfnamefont {W.}~\bibnamefont {Schmidt}}, \ and\
  \bibinfo {author} {\bibfnamefont {D.~R.~G.}\ \bibnamefont {Schleicher}},\
  }\href {\doibase 10.1063/1.4954304} {\bibfield  {journal} {\bibinfo
  {journal} {Physics of Plasmas}\ }\textbf {\bibinfo {volume} {23}},\ \bibinfo
  {eid} {062317} (\bibinfo {year} {2016}),\ 10.1063/1.4954304}\BibitemShut
  {NoStop}%
\bibitem [{\citenamefont {{Smagorinsky}}(1963)}]{Smagorinsky1963}%
  \BibitemOpen
  \bibfield  {author} {\bibinfo {author} {\bibfnamefont {J.}~\bibnamefont
  {{Smagorinsky}}},\ }\href {\doibase
  10.1175/1520-0493(1963)091<0099:GCEWTP>2.3.CO;2} {\bibfield  {journal}
  {\bibinfo  {journal} {Monthly Weather Review}\ }\textbf {\bibinfo {volume}
  {91}},\ \bibinfo {pages} {99} (\bibinfo {year} {1963})}\BibitemShut {NoStop}%
\bibitem [{\citenamefont {Vreman}\ \emph {et~al.}(1994)\citenamefont {Vreman},
  \citenamefont {Geurts},\ and\ \citenamefont {Kuerten}}]{FLM:353319}%
  \BibitemOpen
  \bibfield  {author} {\bibinfo {author} {\bibfnamefont {B.}~\bibnamefont
  {Vreman}}, \bibinfo {author} {\bibfnamefont {B.}~\bibnamefont {Geurts}}, \
  and\ \bibinfo {author} {\bibfnamefont {H.}~\bibnamefont {Kuerten}},\ }\href
  {\doibase 10.1017/S0022112094003745} {\bibfield  {journal} {\bibinfo
  {journal} {Journal of Fluid Mechanics}\ }\textbf {\bibinfo {volume} {278}},\
  \bibinfo {pages} {351} (\bibinfo {year} {1994})}\BibitemShut {NoStop}%
\bibitem [{\citenamefont {Bardina}\ \emph {et~al.}(1980)\citenamefont
  {Bardina}, \citenamefont {Ferziger},\ and\ \citenamefont
  {Reynolds}}]{BARDINA1980}%
  \BibitemOpen
  \bibfield  {author} {\bibinfo {author} {\bibfnamefont {J.}~\bibnamefont
  {Bardina}}, \bibinfo {author} {\bibfnamefont {J.}~\bibnamefont {Ferziger}}, \
  and\ \bibinfo {author} {\bibfnamefont {W.}~\bibnamefont {Reynolds}}\
  }(\bibinfo  {publisher} {American Institute of Aeronautics and
  Astronautics},\ \bibinfo {year} {1980})\BibitemShut {NoStop}%
\bibitem [{\citenamefont {Yeo}(1987)}]{Yeo87}%
  \BibitemOpen
  \bibfield  {author} {\bibinfo {author} {\bibfnamefont {W.~K.}\ \bibnamefont
  {Yeo}},\ }\emph {\bibinfo {title} {{A generalized high pass/low pass
  averaging procedure for deriving and solving turbulent flow equations}}},\
  \href@noop {} {Ph.D. thesis},\ \bibinfo  {school} {The Ohio State University}
  (\bibinfo {year} {1987})\BibitemShut {NoStop}%
\bibitem [{\citenamefont {Bryan}\ \emph {et~al.}(2014)\citenamefont {Bryan},
  \citenamefont {Norman}, \citenamefont {O'Shea}, \citenamefont {Abel},
  \citenamefont {Wise}, \citenamefont {Turk}, \citenamefont {Reynolds},
  \citenamefont {Collins}, \citenamefont {Wang}, \citenamefont {Skillman},
  \citenamefont {Smith}, \citenamefont {Harkness}, \citenamefont {Bordner},
  \citenamefont {hoon Kim}, \citenamefont {Kuhlen}, \citenamefont {Xu},
  \citenamefont {Goldbaum}, \citenamefont {Hummels}, \citenamefont {Kritsuk},
  \citenamefont {Tasker}, \citenamefont {Skory}, \citenamefont {Simpson},
  \citenamefont {Hahn}, \citenamefont {Oishi}, \citenamefont {So},
  \citenamefont {Zhao}, \citenamefont {Cen}, \citenamefont {Li},\ and\
  \citenamefont {Collaboration}}]{Enzo2013}%
  \BibitemOpen
  \bibfield  {author} {\bibinfo {author} {\bibfnamefont {G.~L.}\ \bibnamefont
  {Bryan}}, \bibinfo {author} {\bibfnamefont {M.~L.}\ \bibnamefont {Norman}},
  \bibinfo {author} {\bibfnamefont {B.~W.}\ \bibnamefont {O'Shea}}, \bibinfo
  {author} {\bibfnamefont {T.}~\bibnamefont {Abel}}, \bibinfo {author}
  {\bibfnamefont {J.~H.}\ \bibnamefont {Wise}}, \bibinfo {author}
  {\bibfnamefont {M.~J.}\ \bibnamefont {Turk}}, \bibinfo {author}
  {\bibfnamefont {D.~R.}\ \bibnamefont {Reynolds}}, \bibinfo {author}
  {\bibfnamefont {D.~C.}\ \bibnamefont {Collins}}, \bibinfo {author}
  {\bibfnamefont {P.}~\bibnamefont {Wang}}, \bibinfo {author} {\bibfnamefont
  {S.~W.}\ \bibnamefont {Skillman}}, \bibinfo {author} {\bibfnamefont
  {B.}~\bibnamefont {Smith}}, \bibinfo {author} {\bibfnamefont {R.~P.}\
  \bibnamefont {Harkness}}, \bibinfo {author} {\bibfnamefont {J.}~\bibnamefont
  {Bordner}}, \bibinfo {author} {\bibfnamefont {J.}~\bibnamefont {hoon Kim}},
  \bibinfo {author} {\bibfnamefont {M.}~\bibnamefont {Kuhlen}}, \bibinfo
  {author} {\bibfnamefont {H.}~\bibnamefont {Xu}}, \bibinfo {author}
  {\bibfnamefont {N.}~\bibnamefont {Goldbaum}}, \bibinfo {author}
  {\bibfnamefont {C.}~\bibnamefont {Hummels}}, \bibinfo {author} {\bibfnamefont
  {A.~G.}\ \bibnamefont {Kritsuk}}, \bibinfo {author} {\bibfnamefont
  {E.}~\bibnamefont {Tasker}}, \bibinfo {author} {\bibfnamefont
  {S.}~\bibnamefont {Skory}}, \bibinfo {author} {\bibfnamefont {C.~M.}\
  \bibnamefont {Simpson}}, \bibinfo {author} {\bibfnamefont {O.}~\bibnamefont
  {Hahn}}, \bibinfo {author} {\bibfnamefont {J.~S.}\ \bibnamefont {Oishi}},
  \bibinfo {author} {\bibfnamefont {G.~C.}\ \bibnamefont {So}}, \bibinfo
  {author} {\bibfnamefont {F.}~\bibnamefont {Zhao}}, \bibinfo {author}
  {\bibfnamefont {R.}~\bibnamefont {Cen}}, \bibinfo {author} {\bibfnamefont
  {Y.}~\bibnamefont {Li}}, \ and\ \bibinfo {author} {\bibfnamefont {T.~E.}\
  \bibnamefont {Collaboration}},\ }\href
  {http://stacks.iop.org/0067-0049/211/i=2/a=19} {\bibfield  {journal}
  {\bibinfo  {journal} {The Astrophysical Journal Supplement Series}\ }\textbf
  {\bibinfo {volume} {211}},\ \bibinfo {pages} {19} (\bibinfo {year}
  {2014})}\BibitemShut {NoStop}%
\bibitem [{\citenamefont {Vasilyev}\ \emph {et~al.}(1998)\citenamefont
  {Vasilyev}, \citenamefont {Lund},\ and\ \citenamefont
  {Moin}}]{Vasilyev199882}%
  \BibitemOpen
  \bibfield  {author} {\bibinfo {author} {\bibfnamefont {O.~V.}\ \bibnamefont
  {Vasilyev}}, \bibinfo {author} {\bibfnamefont {T.~S.}\ \bibnamefont {Lund}},
  \ and\ \bibinfo {author} {\bibfnamefont {P.}~\bibnamefont {Moin}},\ }\href
  {\doibase 10.1006/jcph.1998.6060} {\bibfield  {journal} {\bibinfo  {journal}
  {Journal of Computational Physics}\ }\textbf {\bibinfo {volume} {146}},\
  \bibinfo {pages} {82 } (\bibinfo {year} {1998})}\BibitemShut {NoStop}%
\bibitem [{\citenamefont {Sagaut}\ and\ \citenamefont
  {Grohens}(1999)}]{FLD:FLD914}%
  \BibitemOpen
  \bibfield  {author} {\bibinfo {author} {\bibfnamefont {P.}~\bibnamefont
  {Sagaut}}\ and\ \bibinfo {author} {\bibfnamefont {R.}~\bibnamefont
  {Grohens}},\ }\href {\doibase
  10.1002/(SICI)1097-0363(19991230)31:8<1195::AID-FLD914>3.0.CO;2-H} {\bibfield
   {journal} {\bibinfo  {journal} {International Journal for Numerical Methods
  in Fluids}\ }\textbf {\bibinfo {volume} {31}},\ \bibinfo {pages} {1195}
  (\bibinfo {year} {1999})}\BibitemShut {NoStop}%
\bibitem [{\citenamefont {{Schmidt}}\ \emph {et~al.}(2009)\citenamefont
  {{Schmidt}}, \citenamefont {{Federrath}}, \citenamefont {{Hupp}},
  \citenamefont {{Kern}},\ and\ \citenamefont {{Niemeyer}}}]{Schmidt2009}%
  \BibitemOpen
  \bibfield  {author} {\bibinfo {author} {\bibfnamefont {W.}~\bibnamefont
  {{Schmidt}}}, \bibinfo {author} {\bibfnamefont {C.}~\bibnamefont
  {{Federrath}}}, \bibinfo {author} {\bibfnamefont {M.}~\bibnamefont {{Hupp}}},
  \bibinfo {author} {\bibfnamefont {S.}~\bibnamefont {{Kern}}}, \ and\ \bibinfo
  {author} {\bibfnamefont {J.~C.}\ \bibnamefont {{Niemeyer}}},\ }\href
  {\doibase 10.1051/0004-6361:200809967} {\bibfield  {journal} {\bibinfo
  {journal} {Astronomy \& Astrophysics}\ }\textbf {\bibinfo {volume} {494}},\
  \bibinfo {pages} {127} (\bibinfo {year} {2009})}\BibitemShut {NoStop}%
\bibitem [{\citenamefont {Zhou}\ \emph {et~al.}(2014)\citenamefont {Zhou},
  \citenamefont {Grinstein}, \citenamefont {Wachtor},\ and\ \citenamefont
  {Haines}}]{PhysRevE.89.013303}%
  \BibitemOpen
  \bibfield  {author} {\bibinfo {author} {\bibfnamefont {Y.}~\bibnamefont
  {Zhou}}, \bibinfo {author} {\bibfnamefont {F.~F.}\ \bibnamefont {Grinstein}},
  \bibinfo {author} {\bibfnamefont {A.~J.}\ \bibnamefont {Wachtor}}, \ and\
  \bibinfo {author} {\bibfnamefont {B.~M.}\ \bibnamefont {Haines}},\ }\href
  {\doibase 10.1103/PhysRevE.89.013303} {\bibfield  {journal} {\bibinfo
  {journal} {Phys. Rev. E}\ }\textbf {\bibinfo {volume} {89}},\ \bibinfo
  {pages} {013303} (\bibinfo {year} {2014})}\BibitemShut {NoStop}%
\bibitem [{\citenamefont {Shebalin}\ \emph {et~al.}(2009)\citenamefont
  {Shebalin}, \citenamefont {Matthaeus},\ and\ \citenamefont
  {Montgomery}}]{shebalin1983}%
  \BibitemOpen
  \bibfield  {author} {\bibinfo {author} {\bibfnamefont {J.~V.}\ \bibnamefont
  {Shebalin}}, \bibinfo {author} {\bibfnamefont {W.~H.}\ \bibnamefont
  {Matthaeus}}, \ and\ \bibinfo {author} {\bibfnamefont {D.}~\bibnamefont
  {Montgomery}},\ }\href {\doibase 10.1017/S0022377800000933} {\bibfield
  {journal} {\bibinfo  {journal} {Journal of Plasma Physics}\ }\textbf
  {\bibinfo {volume} {29}},\ \bibinfo {pages} {525} (\bibinfo {year}
  {2009})}\BibitemShut {NoStop}%
\bibitem [{\citenamefont {Oughton}\ \emph {et~al.}(1994)\citenamefont
  {Oughton}, \citenamefont {Priest},\ and\ \citenamefont
  {Matthaeus}}]{oughton1994}%
  \BibitemOpen
  \bibfield  {author} {\bibinfo {author} {\bibfnamefont {S.}~\bibnamefont
  {Oughton}}, \bibinfo {author} {\bibfnamefont {E.~R.}\ \bibnamefont {Priest}},
  \ and\ \bibinfo {author} {\bibfnamefont {W.~H.}\ \bibnamefont {Matthaeus}},\
  }\href {\doibase 10.1017/S0022112094002867} {\bibfield  {journal} {\bibinfo
  {journal} {Journal of Fluid Mechanics}\ }\textbf {\bibinfo {volume} {280}},\
  \bibinfo {pages} {95} (\bibinfo {year} {1994})}\BibitemShut {NoStop}%
\bibitem [{\citenamefont {Biskamp}(2003)}]{biskamp}%
  \BibitemOpen
  \bibfield  {author} {\bibinfo {author} {\bibfnamefont {D.}~\bibnamefont
  {Biskamp}},\ }\href@noop {} {\emph {\bibinfo {title} {Magnetohydrodynamic
  Turbulence, by Dieter Biskamp, pp.~310.~ISBN 0521810116.~Cambridge, UK:
  Cambridge University Press, September 2003.}}}\ (\bibinfo  {publisher}
  {Cambrdige University Press},\ \bibinfo {year} {2003})\BibitemShut {NoStop}%
\bibitem [{\citenamefont {Benzi}\ \emph {et~al.}(1993)\citenamefont {Benzi},
  \citenamefont {Ciliberto}, \citenamefont {Tripiccione}, \citenamefont
  {Baudet}, \citenamefont {Massaioli},\ and\ \citenamefont
  {Succi}}]{PhysRevE.48.R29}%
  \BibitemOpen
  \bibfield  {author} {\bibinfo {author} {\bibfnamefont {R.}~\bibnamefont
  {Benzi}}, \bibinfo {author} {\bibfnamefont {S.}~\bibnamefont {Ciliberto}},
  \bibinfo {author} {\bibfnamefont {R.}~\bibnamefont {Tripiccione}}, \bibinfo
  {author} {\bibfnamefont {C.}~\bibnamefont {Baudet}}, \bibinfo {author}
  {\bibfnamefont {F.}~\bibnamefont {Massaioli}}, \ and\ \bibinfo {author}
  {\bibfnamefont {S.}~\bibnamefont {Succi}},\ }\href {\doibase
  10.1103/PhysRevE.48.R29} {\bibfield  {journal} {\bibinfo  {journal} {Phys.
  Rev. E}\ }\textbf {\bibinfo {volume} {48}},\ \bibinfo {pages} {R29} (\bibinfo
  {year} {1993})}\BibitemShut {NoStop}%
\bibitem [{\citenamefont {Newville}\ \emph {et~al.}(2014)\citenamefont
  {Newville}, \citenamefont {Stensitzki}, \citenamefont {Allen},\ and\
  \citenamefont {Ingargiola}}]{newville_2014_11813}%
  \BibitemOpen
  \bibfield  {author} {\bibinfo {author} {\bibfnamefont {M.}~\bibnamefont
  {Newville}}, \bibinfo {author} {\bibfnamefont {T.}~\bibnamefont
  {Stensitzki}}, \bibinfo {author} {\bibfnamefont {D.~B.}\ \bibnamefont
  {Allen}}, \ and\ \bibinfo {author} {\bibfnamefont {A.}~\bibnamefont
  {Ingargiola}},\ }\href {\doibase 10.5281/zenodo.11813} {\enquote {\bibinfo
  {title} {{LMFIT: Non-Linear Least-Square Minimization and Curve-Fitting for
  Python}},}\ } (\bibinfo {year} {2014})\BibitemShut {NoStop}%
\bibitem [{\citenamefont {She}\ and\ \citenamefont
  {Leveque}(1994)}]{PhysRevLett.72.336}%
  \BibitemOpen
  \bibfield  {author} {\bibinfo {author} {\bibfnamefont {Z.-S.}\ \bibnamefont
  {She}}\ and\ \bibinfo {author} {\bibfnamefont {E.}~\bibnamefont {Leveque}},\
  }\href {\doibase 10.1103/PhysRevLett.72.336} {\bibfield  {journal} {\bibinfo
  {journal} {Phys. Rev. Lett.}\ }\textbf {\bibinfo {volume} {72}},\ \bibinfo
  {pages} {336} (\bibinfo {year} {1994})}\BibitemShut {NoStop}%
\bibitem [{\citenamefont {Ghosal}(1996)}]{GHOSAL1996187}%
  \BibitemOpen
  \bibfield  {author} {\bibinfo {author} {\bibfnamefont {S.}~\bibnamefont
  {Ghosal}},\ }\href {\doibase 10.1006/jcph.1996.0088} {\bibfield  {journal}
  {\bibinfo  {journal} {Journal of Computational Physics}\ }\textbf {\bibinfo
  {volume} {125}},\ \bibinfo {pages} {187 } (\bibinfo {year}
  {1996})}\BibitemShut {NoStop}%
\bibitem [{\citenamefont {Chow}\ and\ \citenamefont
  {Moin}(2003)}]{Chow2003366}%
  \BibitemOpen
  \bibfield  {author} {\bibinfo {author} {\bibfnamefont {F.~K.}\ \bibnamefont
  {Chow}}\ and\ \bibinfo {author} {\bibfnamefont {P.}~\bibnamefont {Moin}},\
  }\href {\doibase 10.1016/S0021-9991(02)00020-7} {\bibfield  {journal}
  {\bibinfo  {journal} {Journal of Computational Physics}\ }\textbf {\bibinfo
  {volume} {184}},\ \bibinfo {pages} {366 } (\bibinfo {year}
  {2003})}\BibitemShut {NoStop}%
\bibitem [{\citenamefont {Garnier}\ \emph {et~al.}(1999)\citenamefont
  {Garnier}, \citenamefont {Mossi}, \citenamefont {Sagaut}, \citenamefont
  {Comte},\ and\ \citenamefont {Deville}}]{Garnier1999273}%
  \BibitemOpen
  \bibfield  {author} {\bibinfo {author} {\bibfnamefont {E.}~\bibnamefont
  {Garnier}}, \bibinfo {author} {\bibfnamefont {M.}~\bibnamefont {Mossi}},
  \bibinfo {author} {\bibfnamefont {P.}~\bibnamefont {Sagaut}}, \bibinfo
  {author} {\bibfnamefont {P.}~\bibnamefont {Comte}}, \ and\ \bibinfo {author}
  {\bibfnamefont {M.}~\bibnamefont {Deville}},\ }\href {\doibase
  10.1006/jcph.1999.6268} {\bibfield  {journal} {\bibinfo  {journal} {Journal
  of Computational Physics}\ }\textbf {\bibinfo {volume} {153}},\ \bibinfo
  {pages} {273 } (\bibinfo {year} {1999})}\BibitemShut {NoStop}%
\bibitem [{\citenamefont {Winckelmans}\ \emph {et~al.}(2001)\citenamefont
  {Winckelmans}, \citenamefont {Wray}, \citenamefont {Vasilyev},\ and\
  \citenamefont
  {Jeanmart}}]{:/content/aip/journal/pof2/13/5/10.1063/1.1360192}%
  \BibitemOpen
  \bibfield  {author} {\bibinfo {author} {\bibfnamefont {G.~S.}\ \bibnamefont
  {Winckelmans}}, \bibinfo {author} {\bibfnamefont {A.~A.}\ \bibnamefont
  {Wray}}, \bibinfo {author} {\bibfnamefont {O.~V.}\ \bibnamefont {Vasilyev}},
  \ and\ \bibinfo {author} {\bibfnamefont {H.}~\bibnamefont {Jeanmart}},\
  }\href {\doibase 10.1063/1.1360192} {\bibfield  {journal} {\bibinfo
  {journal} {Physics of Fluids}\ }\textbf {\bibinfo {volume} {13}},\ \bibinfo
  {pages} {1385} (\bibinfo {year} {2001})}\BibitemShut {NoStop}%
\bibitem [{\citenamefont {Chernyshov}\ \emph {et~al.}(2012)\citenamefont
  {Chernyshov}, \citenamefont {Karelsky},\ and\ \citenamefont
  {Petrosyan}}]{Chernyshov2012}%
  \BibitemOpen
  \bibfield  {author} {\bibinfo {author} {\bibfnamefont {A.}~\bibnamefont
  {Chernyshov}}, \bibinfo {author} {\bibfnamefont {K.}~\bibnamefont
  {Karelsky}}, \ and\ \bibinfo {author} {\bibfnamefont {A.}~\bibnamefont
  {Petrosyan}},\ }\href {\doibase 10.1007/s10494-012-9408-x} {\bibfield
  {journal} {\bibinfo  {journal} {Flow, Turbulence and Combustion}\ }\textbf
  {\bibinfo {volume} {89}},\ \bibinfo {pages} {563} (\bibinfo {year}
  {2012})}\BibitemShut {NoStop}%
\bibitem [{\citenamefont {Zhou}\ and\ \citenamefont
  {Oughton}(2011)}]{Zhou2011}%
  \BibitemOpen
  \bibfield  {author} {\bibinfo {author} {\bibfnamefont {Y.}~\bibnamefont
  {Zhou}}\ and\ \bibinfo {author} {\bibfnamefont {S.}~\bibnamefont {Oughton}},\
  }\href {\doibase 10.1063/1.3606473} {\bibfield  {journal} {\bibinfo
  {journal} {Physics of Plasmas}\ }\textbf {\bibinfo {volume} {18}},\ \bibinfo
  {pages} {072304} (\bibinfo {year} {2011})},\ \Eprint
  {http://arxiv.org/abs/http://aip.scitation.org/doi/pdf/10.1063/1.3606473}
  {http://aip.scitation.org/doi/pdf/10.1063/1.3606473} \BibitemShut {NoStop}%
\bibitem [{\citenamefont {Zhou}(2010)}]{Zhou20101}%
  \BibitemOpen
  \bibfield  {author} {\bibinfo {author} {\bibfnamefont {Y.}~\bibnamefont
  {Zhou}},\ }\href {\doibase http://dx.doi.org/10.1016/j.physrep.2009.04.004}
  {\bibfield  {journal} {\bibinfo  {journal} {Physics Reports}\ }\textbf
  {\bibinfo {volume} {488}},\ \bibinfo {pages} {1 } (\bibinfo {year}
  {2010})}\BibitemShut {NoStop}%
\bibitem [{\citenamefont {Zhou}\ \emph {et~al.}(2004)\citenamefont {Zhou},
  \citenamefont {Matthaeus},\ and\ \citenamefont {Dmitruk}}]{Zhou2004}%
  \BibitemOpen
  \bibfield  {author} {\bibinfo {author} {\bibfnamefont {Y.}~\bibnamefont
  {Zhou}}, \bibinfo {author} {\bibfnamefont {W.~H.}\ \bibnamefont {Matthaeus}},
  \ and\ \bibinfo {author} {\bibfnamefont {P.}~\bibnamefont {Dmitruk}},\ }\href
  {\doibase 10.1103/RevModPhys.76.1015} {\bibfield  {journal} {\bibinfo
  {journal} {Rev. Mod. Phys.}\ }\textbf {\bibinfo {volume} {76}},\ \bibinfo
  {pages} {1015} (\bibinfo {year} {2004})}\BibitemShut {NoStop}%
\end{thebibliography}
\end{document}